\documentclass[intlimits,twoside,a4paper]{article}

\usepackage{amsmath,amssymb} 
\usepackage{graphicx}        
\usepackage{scalerel}        
\usepackage{tikz}            
\usepackage{bm}              
\usepackage[T2A,T1]{fontenc} 
\usepackage{multirow,array}
\usepackage{graphicx}
\usepackage{bm}
\usepackage{float}
\usepackage{color}\usepackage{subfigure}
\usepackage{verbatim}
\usepackage{array}
\usepackage{booktabs}
\usepackage{soul}
\usepackage[cp1251]{inputenc}

\usepackage[eqsecnum]{cmpj3}

\issue{2026}{29}{1}{13601}
\doinumber{10.5488/CMP.29.13601}

\title[Properties of aqueous perchloric acid]
{
Molecular dynamics study of perchloric acid using the extended Madrid-2019 force field
}

\author[M. Cruz-S\'anchez, S. Blazquez, C. Vega, V. M. Trejos]{
M. Cruz-S\'anchez\orcid{0009-0005-0407-6496}\refaddr{label1},
S. Blazquez\orcid{0000-0002-6218-3880}\refaddr{label2},
C. Vega\orcid{0000-0002-2417-9645}\refaddr{label2},
V. M. Trejos\orcid{0000-0002-1136-3806}\refaddr{label1}
\thanks{Corresponding author: \email{vtrejos@izt.uam.mx}}
}
\addresses{
\addr{label1} Departamento de Qu\'imica, Universidad Aut\'onoma Metropolitana-Iztapalapa,
Av. San Rafael Atlixco 186, Col.~Vicentina, 09340, Ciudad de M\'exico, M\'exico
\addr{label2} Departamento de Qu\'imica F\'isica, Universidad Complutense de Madrid, 28040 Madrid, Spain
}

\date{Received 04 November 2025; revised 17 December 2025; accepted 03 January 2026; published 30 March 2026}
\begin{document}

\maketitle

\begin{abstract}

Perchloric acid  (HClO$_4$) is widely used to prepare perchlorate salts with applications in propellants,
industry, environmental chemistry, and biology.
In this work, we used the intermolecular parameters from the extended Madrid-2019
force field for the perchlorate anion (ClO$_4^-$) and the oxonium cation (H$_3$O$^+$) together with
TIP4P/2005 water to model perchloric acid solutions.
The force field uses scaled charges of $\pm0.85e$ for monovalent ions and has been
widely applied for aqueous ionic systems.
We used the model to predict thermodynamic properties [densities and temperatures of maximum in density (TMD)],
structural features (ion-water correlations: ion-hydrogen and ion-oxygen), and transport properties
(self-diffusion coefficients and viscosity) of perchloric acid solutions at several concentrations.
Experimental densities are predicted in excellent agreement up to 10~$m$.
We also performed molecular simulations over a wide range of temperatures in order to determine the
TMD of perchloric acid at different molalities.
Predicted viscosities at 298.15 K and 1 bar are in good agreement with experimental data
for concentrations below 4 $m$.
Results are discussed in terms of model strengths and limitations.

\end{abstract}

\section{Introduction}

Perchloric acid (HClO$_4$) is an oxoacid of chlorine that is typically encountered as an aqueous solution.
Its volumetric, rheological, and transport properties are of special importance from both fundamental
and applied perspectives.
Accurate data for density, viscosity, and diffusion coefficients across wide concentration and
temperature ranges are essential for thermodynamic modelling, process design, and validation
of molecular simulations.
Perchloric acid is stronger than nitric and sulfuric acids, is corrosive to tissues and metals,
and is widely used in the preparation of perchlorate salts
(e.g., ammonium perchlorate), which are important components of rocket propellants.
The strong acidity of HClO$_4$ arises from the high resonance stabilization of the
perchlorate anion ClO$_4^-$.
Structurally, perchloric acid consists of a central chlorine atom bonded to four oxygen atoms,
one of which is also covalently bound to a hydrogen atom. The perchlorate anion is well described
by a tetrahedral arrangement of oxygens around chlorine.

Classical experimental studies provide key reference data
for density-composition relations at high concentrations.
Smith and co-workers~\cite{smith1931} compiled high-precision density tables for 65--75 wt$\%$
HClO$_4$ at 25$^\circ$C using oxonium perchlorate as a standard. 
Markham~\cite{markham1941} reported densities at 25$^\circ$C up to
65~wt\% HClO$_4$, and Brickwedde~\cite{brickwedde1935} extended measurements
of density, viscosity, and electrical resistivity from 0--70 wt$\%$ over temperatures range
$-60$ to $+ 75^\circ$C.
Ionic transport measurements by Malhotra and Woolf~\cite{Malhotra1993} provided intradiffusion
coefficients for tritiated water and perchlorate ion at 25$^\circ$C
over a wide concentration range (0--7.349 mol/L).

The increasing and continued use of perchloric acid solutions in laboratory has prompted a 
detailed experimental characterization of many physicochemical properties.
From a theoretical and computational point of view, it is therefore important to develop
reliable molecular models that can accurately predict thermodynamic, interfacial, and dynamic properties
and that provide microscopic insight into the behavior of these mixtures.
To the best of our knowledge, comprehensive molecular simulation studies of perchloric acid in aqueous
solution remain limited, motivating the need for systematic computational investigations.
In this context, we employ the intermolecular parameters of the force field developed for the perchlorate
anion (ClO$_4^-$)~\cite{Blazquez2025perchlorate} and the oxonium (also called hydronium in the literature) cation
(H$_3$O$^+$)~\cite{Gamez2025_H3O} to construct a molecular model of perchloric acid (HClO$_4$),
assigning a total scaled charge of $q_{\text{scaled}} = \pm 0.85 e$.
As perchloric acid is a strong acid, we shall assume complete dissociation in aqueous solution for each ion, 
such that the system has hydronium and perchlorate ions in water after the acid disolves in water. 
The force field parameters for both ions (hydronium and perchlorate) was taken from the Madrid-2019 force field. The Madrid-2019 force field is a non-polarizable force field 
for electrolytes in water. The two key ingredients are: TIP4P/2005 model for water \cite{Vega2005} and the use of scaled charges for the ions. 
The use of scaled charges for ions requires some motivation.  Leontyev and  Stuchebrukhov  
     pointed out that non-polarizable models of 
     water lack the electronic contribution to polarizability \cite{leontyev11}.  
    At high frequencies, where the nuclei cannot follow the oscillations of the electric field,  
    the dielectric constant of water is around 1.78. A model lacking the electronic contribution should have a charge of $1/\sqrt{1.78}=0.75$ (scaled charges are given in electron units). 
    Kann and Skinner pointed out \cite{kan:jcp14} that to reproduce the Debye-Huckel  law, a model of water with a certain value of $\varepsilon$ (at room $T$ and $P$) for the 
    dielectric constant should use charges with the value $\sqrt{(\varepsilon/78)}$.  In the particular case 
    of TIP4P/2005, this leads to a charge of 0.85 and this is the choice of the Madrid-2019 force field. Vega \cite{vegamp15} suggested that different charges should be used to describe the potential energy (PES, scaled charges) and dipole moment surfaces (DMS, formal charges). This idea has been successfully applied to predict  dielectric constants \cite{aragones2011dielectric}  and electrical conductivities~\cite{blazquez-conductivities}.
        Scaled charges are often used in the community performing simulations of ionic-liquids \cite{C2CP23329K} as they are thought to effectively account for polarizability and charge transfer \cite{doi:10.1063/1.4736851}.
       The scaling of the charges is gaining popularity as has been adopted by 
       many authors \cite{dub:jcpb17,jungwirth_2025,fue:jpc16,Habibi2022,predota-ions-neutron,laage_scaled,le2020molecular}. Besides in the particular case of the perchlorate
       anion it has been shown that using formal charges~\cite{ClO4_formal_charges} 
       ({i.e.,} 1 for the charge of the anion) lead to 
    bad predictions of the densities for NaClO$_4$ solutions \cite{Blazquez2025perchlorate}.

Using this model, we investigate viscosities, self-diffusion coefficients,
temperatures of maximum density (TMD), and structural features.
The implications of these results for experimental comparison and future modelling applications are also
discussed.

The manuscript is organized as follows. In section~\ref{section1}, we describe the intermolecular
pair potential, simulation details, and the mathematical expressions used to compute the dynamic
properties of aqueous perchloric acid solutions.
Section~\ref{results} presents the main results concerning thermodynamic properties
(bulk densities and temperatures of maximum density), 
structural properties (radial distribution functions), and
dynamic properties (self-diffusion coefficients and shear viscosities).
Finally, section~\ref{Conclusions} summarizes the principal findings and provides the concluding remarks.

\section{Simulation details}\label{section1}

\subsection{Intermolecular pair potential}

In the present work, we consider a system composed of perchloric acid and water.
The intermolecular pair potential $u(r_{ij})$ between two molecules $i$ and $j$,
which depends on the center-center distance and molecular orientations, is described as
the sum of a Lennard-Jones (LJ) potential and a coulombic interactions,
\begin{equation}
    u(r_{ij}) = 4 \varepsilon_{ij}
    \left[    \left(  \frac{ \sigma_{ij} }{ r_{ij} }  \right)^{12}
    +
    \left(   \frac{ \sigma_{ij} }{ r_{ij} }   \right)^{6}
    \right] +
    \frac{ 1 }{ 4 \piup \varepsilon_0  } \frac{ q_i q_j }{ r_{ij} },
\end{equation}
where $\varepsilon_0$ is the vacuum permittivity,
$\sigma_{ij}$ and $\varepsilon_{ij}$ are the diameter and a well depth of the LJ potential, respectively,
and $q_i$ and $q_j$ are the charges of the atoms $i$ and $j$.
The water model is described by the TIP4P/2005 model~\cite{Vega2005,vega2011}, while the
perchlorate anion (ClO$_4^-$) and oxonium cation (H$_3$O$^+$) are obtained from
further extensions~\cite{Gamez2025_H3O,Blazquez2025perchlorate}
of the Madrid-2019 force field~\cite{zeron2019} model
successfully used for ions in TIP4P/2005 water model
with the scaled value of the charge $q_{\text{scaled}} = \pm0.85e$.
The LJ molecular parameters and the charges are reported in table~\ref{Table1}.

\subsection{Molecular simulations}

Molecular dynamics (MD) simulations were performed using the GROMACS package
(version~4.6.5)~\cite{spoel05,hess08} in both the isothermal-isobaric ($NpT$) and canonical
($NVT$) ensembles.
The leap-frog integration algorithm~\cite{bee:jcp76} was employed with a
time step of 2 fs.
Periodic boundary conditions were applied in all $xyz$ directions.
Temperature and pressure were maintained constant using the Nos\'e--Hoover thermostat~\cite{nose84,hoover85}
and the Parrinello--Rahman barostat~\cite{parrinello81}, respectively, both with coupling time constants of 2 ps.
Long-range energy and pressure corrections were applied to the Lennard-Jones contribution of the potential.
The cutoff radius for both van der Waals and electrostatic interactions was set to 1.0~nm.
Long-range electrostatic interactions were treated using the smooth particle mesh Ewald~(PME) method~\cite{essmann95}.
{HClO$_4$ is treated in this work as a strong acid so that we shall assume that after dissolving in water it forms {i.e.,} H$_3$O$^+$ and ClO$_4^-$.
}
The interaction parameters for the perchlorate anion (ClO$_4^-$) and the oxonium cation
(H$_3$O$^+$) were taken from the extended version of the Madrid-2019 force
field~\cite{Blazquez2025perchlorate,Gamez2025_H3O}, while those for water were obtained from the
TIP4P/2005 model~\cite{Vega2005}.
In the Madrid-2019 force field the ion-water interactions (both for oxonium and perchlorate)  are fitted to reproduce experimental  bulk densities
as a function of electrolyte molality (i.e., mol of acid per kg of water).
In particular, the cross LJ parameters
$\sigma_{{\rm O}_w-{\rm O}_p}$ (where ${\rm O}_w$ and ${\rm O}_p$ denote the oxygen atoms of water and the perchlorate anion,
respectively)  and
$\sigma_{{\rm O}_w-{\rm O}_x}$, $\varepsilon_{{\rm O}_w-{\rm O}_x}$ (where ${\rm O}_x$ denotes the oxygen atom of the oxonium cation).
{In both references~\cite{Blazquez2025perchlorate,Gamez2025_H3O}, the cross LJ parameters
}
were determined by a trial-and-error procedure to optimize the agreement with the experimental data.
For the perchlorate anion, the molecular geometry was taken from reference~\cite{Blazquez2025perchlorate},
with an $\widehat{{\rm O}_p{\rm Cl}_p{\rm O}_p}$ angle of 109.5\textdegree~and a O$_p$--Cl$_p$ distance of 1.43 \AA.
For the oxonium ion, the $\widehat{{\rm HOH}}$ angle is 111.4\textdegree~and the
O--H length is 0.98 $\textup{\AA}$  as reported in reference~\cite{Gamez2025_H3O}.
Implementation details regarding the technique of implementing the geometry of both ions in MD input files
are provided in the Appendixes of references~\cite{Blazquez2025perchlorate,Gamez2025_H3O}.
The optimized potential parameters and charges for the perchlorate and oxonium ions are
reported in tables~\ref{Table1} and~\ref{Table2}. The interactions between the perchlorate
and oxonium cation obey Lorentz--Berthelot combining rules.

\begin{table}[h]
	\centering
	\caption{
		Summary of the force field parameters for the Coulombic (atomic charges, $q_i$) and
		Lennard-Jones ($\sigma_{ii}$, $\varepsilon_{ii}$) parameters
		to the pair potential for the perchlorate anion (ClO$_4^-$) and oxonium cation (H$_3$O$^+$).
		The geometry of the ClO$_4^-$ is defined by a $\widehat{{\rm O}_p{\rm Cl}_p{\rm O}_p}$
		angle of 109.5\textdegree,~and a O$_p$--Cl$_p$ distance of 1.43~$\textup{\AA}$ and a O$_p$--O$_p$ distance of 2.335~$\textup{\AA}$,
		being O$_p$ and Cl$_p$ the oxygen and
		chloride atoms of the perchlorate anion, respectively, as reported in reference~\cite{Blazquez2025perchlorate}.
		In the case of the H$_3$O$^+$ cation the geometry of the ion is defined by a $\widehat{HOH}$ angle
		of 111.4\textdegree~and a O--H distance at 0.98 $\textup{\AA}$ as reported in reference~\cite{Gamez2025_H3O}.
		Subscript $ox$ denote the oxygen atoms from the H$_3$O$^+$ molecule.
		The $\sigma_{ii}$ and $\varepsilon_{ii}$ parameters of the LJ pair potential are presented in units
		of $\textup{\AA}$ and kJ/mol, respectively, while the Coulombic charges are presented in electron
		units ($e$).
		In the case of the perchlorate anion, the model used in this work corresponds to the model
		labeled as $A$ in  reference~\cite{Blazquez2025perchlorate}.
		In both cases, perchlorate anion and oxonium cation, the net charge is $\pm0.85e$ in accordance with the
		Madrid-2019 force field. \\
	}
	\begin{tabular}{
			>{\centering\arraybackslash}m{2.7cm}
			>{\centering\arraybackslash}m{2.7cm}
			>{\centering\arraybackslash}m{2.7cm}
			>{\centering\arraybackslash}m{2.7cm}
		}
		\hline \hline
		\vspace{0.1cm}
		ClO$_4^-$  & $\sigma_{ii}$ ($\textup{\AA}$)   & $\varepsilon_{ii}$  (kJ/mol) &  \textit{q$_i$} ($e$) \\[1ex]
		\hline
		Cl$_p$     & 3.47094                          & 1.10876                     & $+0.35000$                \\
		O$_p$      & 2.95992                          & 0.87864                     & $-0.30000$                \\
		[2ex]
		\hline
		H$_3$O$^+$  & $\sigma_{ii}$ ($\textup{\AA}$)   & $\varepsilon_{ii}$  (kJ/mol) &  \textit{q$_i$} ($e$) \\[1ex]
		\hline
		O$_{ox}$    & 3.10000                          & 0.80000                      & $-0.39442$          \\
		H$_{ox}$    & 0.00000                          & 0.00000                      & $+0.41481$          \\
		[1ex]
		\hline \hline
	\end{tabular}
	\label{Table1}
\end{table}

\begin{table}[h]
	\centering
	\caption{
		Crossed Lennard-Jones ($\sigma_{ij}$) parameters (in \AA) between
		the perchlorate anion (ClO$_4^-$) and oxonium cation (H$_3$O$^+$).
		O$_p$ and Cl$_p$ are the oxygen and chloride atoms of the perchlorate anion, respectively.
		Parameters for TIP4P/2005 water were taken from reference~\cite{Vega2005}.
		LJ parameters for the self- and crossed-interaction between
		chloride atoms of the perchlorate anion are
		taken from the extended Madrid-2019 force field~\cite{Blazquez2025perchlorate,Gamez2025_H3O}.
		In cases where a numerical value is not given, the Lorentz--Berthelot~(LB) combining rules were followed,
		as indicated. O$_w$ stands for the oxygen of water and O$_p$ and Cl$_p$ represent the oxygen
		and chloride atoms of the perchlorate anion, respectively.
		Subscript $ox$ denotes the oxygen atoms from the oxonium cation.
		We indicate whether the LB combining rule is applied.
		The O$_w$ stands for the oxygen of water, O$_{ox}$ represent the oxygen of the
		oxonium cation and O$_p$ and Cl$_p$ represent the oxygen and chloride atoms of the perchlorate anion,
		respectively.
		The LJ  interactions between the perchlorate anion and oxonium cation obey LB combining rules.\\
	}
	\begin{tabular}{
			>{\centering\arraybackslash}m{2cm}
			>{\centering\arraybackslash}m{2cm}
			>{\centering\arraybackslash}m{2cm}
			>{\centering\arraybackslash}m{2cm}
			>{\centering\arraybackslash}m{2cm}
		}
		\hline \hline
		& \multicolumn{2}{ c }{ $\sigma_{ij}$ (\AA)} & \multicolumn{2}{ c }{ $\varepsilon_{ij}$ (kJ/mol) } \\ 
		\cmidrule(lr){2-3}  \cmidrule(lr){4-5}
		\vspace{0.1cm}
		Atom &         Cl$_p$    & O$_p$ &  Cl$_p$    & O$_p$ \\[1ex]
		\hline
		O$_w$      & LB                          & 3.2570       &  LB        & 0.8251           \\
		Cl$_p$     &  ---                         & LB           &  ---        & LB     \\
		[1ex]
		\hline
		& \multicolumn{2}{ c }{ $\sigma_{ij}$ (\AA)} & \multicolumn{2}{ c }{ $\varepsilon_{ij}$ (kJ/mol) } \\ 
		\cmidrule(lr){2-3}  \cmidrule(lr){4-5}
		Atom &        O$_{ox}$     &    H$_{ox}$  &   O$_{ox}$     &    H$_{ox}$ \\[1ex]
		\hline
		O$_w$        & 2.8000      & LB                 & 0.7873  &  LB      \\
		H$_{ox}$     & LB          &  ---               & LB       &  ---      \\
		[1ex]
		\hline \hline
	\end{tabular}
	\label{Table2}
\end{table}

{In the Madrid-2019 both the oxonium and the perchlorate are treated as fully rigid groups. 
In this work the molecular geometry was constrained using the SHAKE algorithm~\cite{ryckaert77}. 
Although the rigid geometry of the oxonium group could be addressed with both LINCS and SHAKE 
algorithms, the tetrahedral geometry of the perchlorate anion can only be forced with SHAKE. 
To keep the tetrahedral geometry requires imposing constraints not only on all Cl--O bond lengths but 
also on the O--O distances. 
The SHAKE algorithm is specifically designed to handle this type of fully rigid, 
multi-constraint molecular geometry. By contrast, LINCS cannot accommodate the complete set 
of O--O constraints needed to preserve the exact tetrahedral structure of ClO$_4^-$.
As documented in Ryckaert {et al.}~\cite{ryckaert77}, SHAKE is the appropriate 
algorithm for enforcing this topology. 
For this reason, SHAKE remains the standard approach for rigid tetrahedral anions in various
MD packages, including LAMMPS and older versions of GROMACS. 
Since recent GROMACS releases no longer support SHAKE for this type of constraint topology, 
we used GROMACS 4.6.5, which still includes native SHAKE support and permits full implementation 
of the rigid-body geometry required for the ClO$_4^-$ anion.
However, for those users that would prefer to use more recent versions of Gromacs 
(where SHAKE is not implemented), they could use the approach proposed in the appendix A of
Blazquez~{et al.}~\cite{Blazquez2025perchlorate} for the perchlorate group where it is shown that it is possible to have an almost rigid (but with a little bit of flexibility) version of the perchlorate group using LINCS and modern version of Gromacs providing almost identical results to those obtained from the entirely rigid model. }
Densities, radial distribution functions (RDFs), and diffusion coefficients were obtained from $NpT$
simulations of systems containing 555 water molecules and the corresponding number of
perchlorate and oxonium ions needed to reach the target molalities; production runs extended for
$\approx50$~ns.

{Perchloric acid (HClO$_4$) reacts with water and is highly dissociated, 
forming H$_3$O$^+$ and ClO$_4^-$ ions. 
Accordingly, our system consists of H$_3$O$^+$ and ClO$_4^-$ ions solvated in water.
When computing the molality, each H$_3$O$^+$ ion must be counted as consuming one water molecule.
Thus, the effective number of water molecules used in the molality calculation is increased 
by the number of H$_3$O$^+$ ions. 
For example, for a system having 10 H$_3$O$^+$ and 10 ClO$_4^-$ units in 555 water molecules, the molality is calculated as
$m = 10 / [(555 + 10) \cdot MW_{\mathrm{wat}} / 1000] = 0.982$, 
where $MW_{\mathrm{wat}}$ is the molecular weight of water. 
Similarly, for a 
system having 7 H$_3$O$^+$ and 7 ClO$_4^-$ units in 555 water molecules,
$m = 7 / [(555 + 7) \cdot MW_{\mathrm{wat}} / 1000] = 0.691$.
Further details regarding the acid dissociation and the definition of the uncorrected molality ($m'$) 
in solutions containing H$_3$O$^+$ species can be found in table I of the Supplementary material 
of reference~\cite{Gamez2025_H3O}.}

\subsection{Dynamics properties}

\subsubsection{Self-diffusion coefficients}

One of the most important target properties used to assess the applicability of force fields 
is the self-diffusion coefficient of each species. 
This property can be obtained from the mean square displacement~(MSD) of the particles. 
In this work, we used the Einstein relation, given by,
\begin{equation}
\label{eq_diffus}
 D_i = \frac{1}{6} \lim_{t \to \infty} \frac{1}{t}
 \big \langle [\mathbf{r}_{i}(t)-\mathbf{r}_{i}(0)]^{2} \big \rangle ,
\end{equation}
where $t$ is the time, $\mathbf{r}_{i}(t)$ and $\mathbf{r}_{i}(0)$ are the position of the $i$-th
particle at time $t$ and at a certain origin of time, respectively,
and the $\langle ... \rangle$ term is the mean square displacement.
The diffusion coefficients were then obtained, avoiding the subdiffusive regime,
from the slope of the plot of the MSD against time.
The so-called values of $D_i$ were subsequently corrected with the hydrodynamic correction of
Yeh and Hummer~\cite{Yehhummer},
\begin{equation}
\label{corectionyeh}
        D_{\text{corr}} = D + 2.837 \frac {k_{\rm B}T}{6\piup \eta L },
\end{equation}
where $ D_{\text{corr}}$ is the corrected diffusion coefficient,
$T$ is the temperature, $k_{\rm B}$ is the Boltzmann constant,
$\eta$ is the simulated viscosity at the studied concentration and $L$ is the length of
the simulation box.

\subsubsection{Shear viscosity}

The shear viscosity was evaluated following the protocol reported in reference~\cite{gon:jcp10}
for rigid models via the self-autocorrelation function of the three off-diagonal components of the pressure tensor $P_{\alpha\beta}(t)$ as formulated in the Green--Kubo formalism,
\begin{equation}
 \eta = \frac {V}{k_{\rm B} T} \int_{0}^{\infty}
 \langle P_{\alpha\beta}(t)\; P_{\alpha\beta}(0)  \rangle\; \rd t,
\end{equation}\\
where $V$ is the system volume.
We first performed $NpT$ simulations of 50 ns to obtain the average equilibrium volume $V$
for each aqueous perchloric acid solution at the desired molality.
Subsequently, $NVT$ simulations of 50 ns were carried out using this average volume,
during which the pressure tensor components $P_{\alpha\beta}(t)$,
were saved on the disk every 2 fs.
The shear viscosity was then computed from the time integral of the autocorrelation function using
the Green--Kubo approach.

\section{Results and discussion}\label{results}

\subsection{Bulk densities}

The primary target property used to validate the extended Madrid-2019 parameters for
perchloric acid was the density.
Remarkably, using only the LJ parameters reported for the perchlorate anion (ClO$_4^-$)
and the oxonium cation (H$_3$O$^+$) was sufficient to construct the perchloric acid model.
The simple use of the LB combining rules for the perchlorate-oxonium interactions was enough for
an accurate description of the densities of the mixture, so that it was not necessary to fit the cross-interactions between the cation and the anion for obtaining good agreement with experiment.

\begin{figure*}
\begin{center}
\includegraphics[height=6cm,clip]{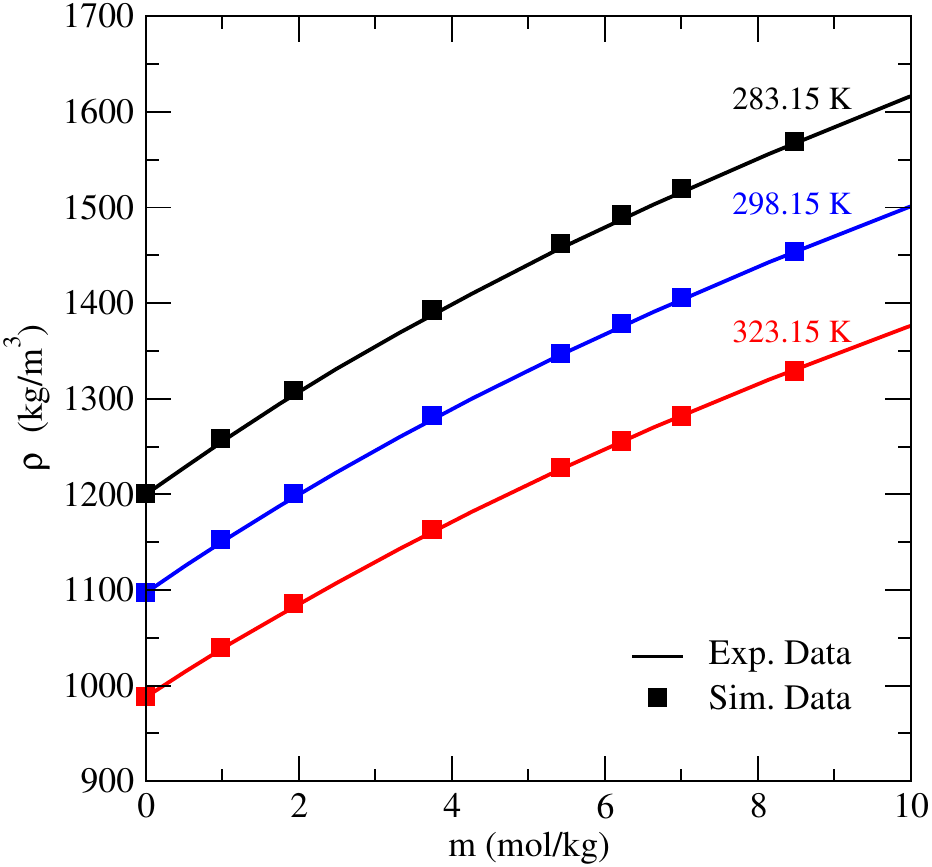}
\end{center}
\caption{(Colour online) 
Density as a function of molality for aqueous solutions of perchloric acid (HClO$_4$) at three
temperatures (namely 298.15, 283.15, and 323.15 K) and 1 bar.
Solid symbols represent the MD results, while the continuous line corresponds
to the experimental data.
Experimental data are taken from reference~\cite{Brickwedde1949properties}.
The MD simulations were performed using the Madrid-2019 force field
($q = \pm0.85e$)~\cite{Blazquez2025perchlorate,Gamez2025_H3O}.
For better visualization, the densities of perchloric acid at 298.15 K and 283.15~K are 
shifted upward by 100 and 200 units, respectively.
}
\label{Fig:1}
\end{figure*}

In figure~\ref{Fig:1}, we present a comparison between
experimental data (continuous line) and our simulation results (solid symbols)
for aqueous solution densities as a function of perchloric acid molality at three different temperatures.
{The purpose of figure~\ref{Fig:1} is to demonstrate that, although the force field parameters for
ClO$_4^-$ and H$_3$O$^+$ were originally developed independently, there was no prior evidence 
that their  combination along with  LB combining rules for the oxonium-perchlorate interactions  would successfully reproduce the experimental densities of perchloric acid
solutions over a wide range of molalities at 298 K. 
Our results show that the Madrid-2019 force field for hydronium and perchlorate ions (along with LB combining rules for the cross interactions between oxonium and perchlorate) indeed yields excellent 
agreement with experimental densities.
As far as we know this is the first time that computer simulations reproduce the experimental densities of 
perchloric  acid in a wide range of concentrations and at different temperatures. 
Moreover, we extended the validation to two additional temperatures, 283.15 K and 323.15 K, 
which were not included in the original parameterization or fitting procedures of the
cited references~\cite{Blazquez2025perchlorate,Gamez2025_H3O}. 
These results demonstrate the temperature transferability and robustness of the model.
Additionally, for 
}
all the investigated temperatures, the force field reproduces experimental densities
accurately up to molalities of about 10~$m$
(a 10 $m$ solution corresponds to a solution with a percentage in weight of about 50$\%$).
The agreement between experiment and simulation is remarkable over the full temperature range considered.
The relative percentage deviations are less than 0.5$\%$ in all the range of molalities
(see table S1 and S2 of the Supplementary Material).
Perchloric acid is highly soluble in water, and no solubility limit has been reported in the literature;
it readily dissociates into hydronium and perchlorate ions.
For reference, commercial perchloric acid is typically supplied as aqueous solutions
containing roughly 60--72$\%$ by weight.

\begin{figure*}
\begin{center}
\includegraphics[height=6.9cm,clip]{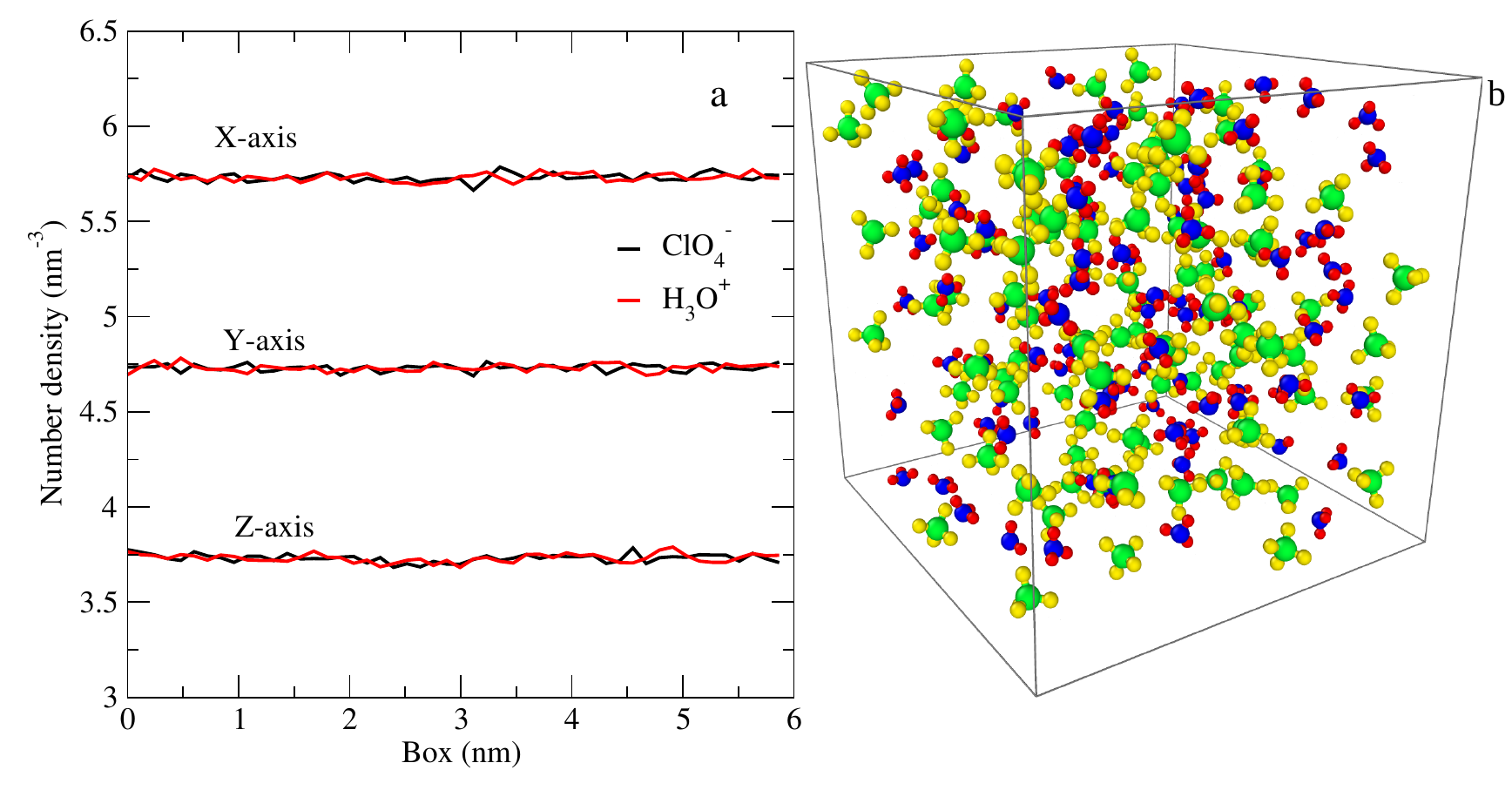}
\end{center}
\caption{(Colour online) 
{Panel (a): 
number-density profiles of ClO$_4^-$ and H$_3$O$^+$ ions as a function of the box length along 
the three Cartesian coordinates ($x$, $y$, and $z$) for the 8.48 $m$ HClO$_4$ solution. 
The profiles are nearly flat and overlap in all directions, indicating that the system remains
homogeneous, with no evidence of aggregation. 
Only small statistical fluctuations ($\approx 1\%$) are observed across the simulation box. 
For clarity, the $y$-axis and $x$-axis values in the number density 
were shifted upward by one and two number density units, respectively.
Panel (b):
representative snapshot of the system at the highest molality studied in this work (8.48 $m$), 
showing the spatial distribution of perchlorate (green and yellow) and hydronium (blue and red) ions. 
}
}
\label{Fig:2new}
\end{figure*}

{Accordingly, for the highest concentration considered (i.e., 8.48~$m$), we also simulated a larger system containing 4440 water molecules 
together with the corresponding number of perchloric acid molecules. 
These simulations were performed for 50 ns in the $NVT$ ensemble to visually confirm the 
absence of precipitation of the acid in the aqueous solution. 
Systems of the same size were additionally used for the calculation of the diffusion coefficient, 
$D$, and the shear viscosity, $\eta$.
Furthermore, in figure~\ref{Fig:2new}, we show a figure consisting of two 
panels: (a) the number-density profiles along the simulation box, and 
(b) a representative snapshot of the system at the highest perchloric acid concentration studied
(i.e., 8.48 $m$).  
Together, these panels provide numerical and visual evidence that no precipitation occurs 
in any of the simulated systems.
}

\newpage
\subsection{Temperature of maximum in density}

\begin{figure*}
\begin{center}
\includegraphics[height=6cm,clip]{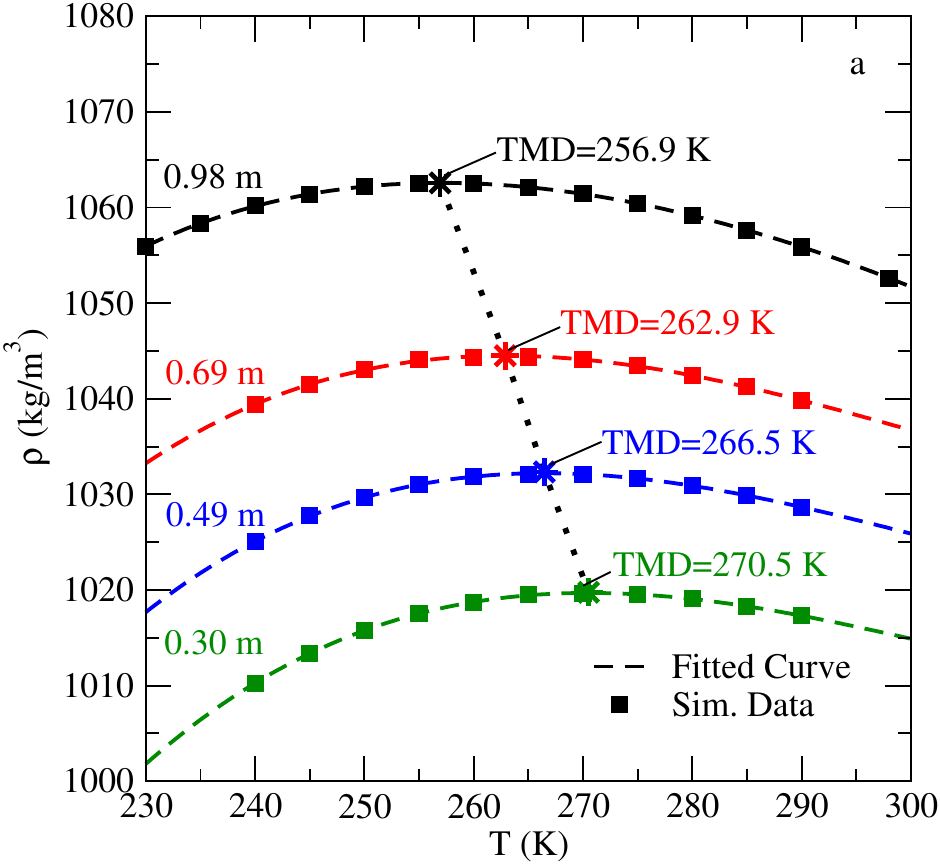}
\includegraphics[height=6cm,clip]{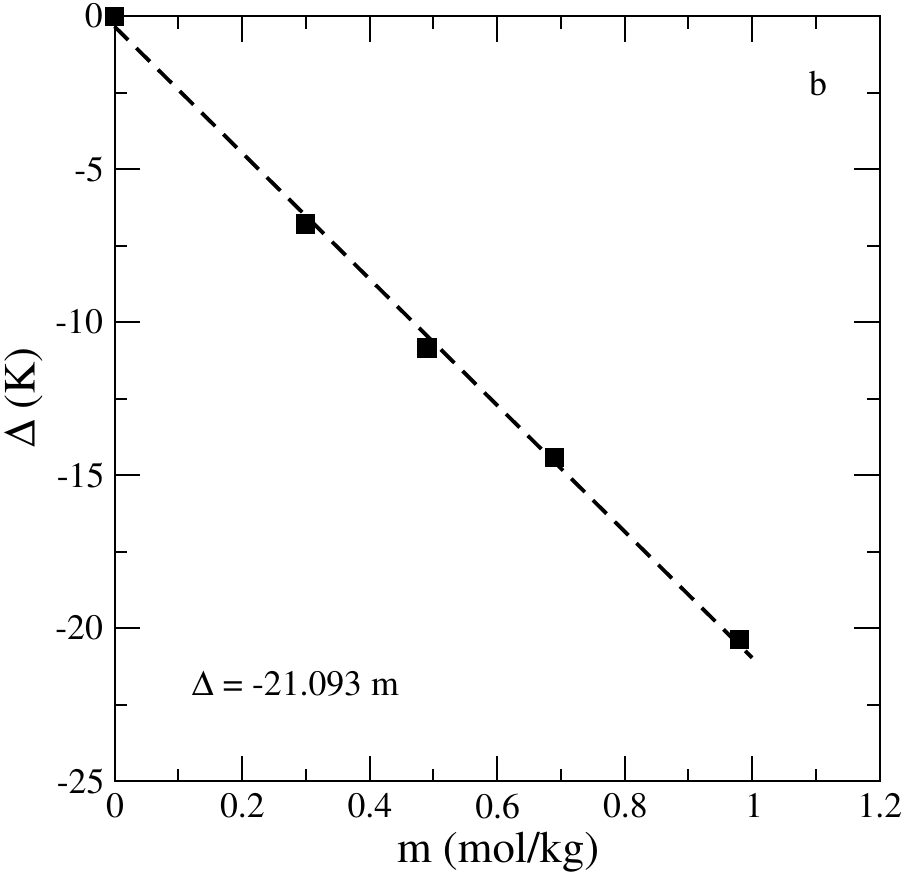}
\end{center}
\caption{(Colour online) 
Panel (a):
density as a function of temperature for aqueous solutions of perchloric acid (HClO$_4$)
at 0.30 $m$, 0.49 $m$, 0.69 $m$, and 0.98 $m$ at 1 bar.
Solid squares denote MD simulation results,
dashed lines represent third-order polynomial fits to the MD data obtained with the
extended Madrid-2019 model~\cite{Blazquez2025perchlorate,Gamez2025_H3O},
and stars indicate the temperature of maximum density.
Panel (b):
variation of $\Delta$ as a function of $m$ for perchloric acid solutions.
The dashed line corresponds to a linear fit to the MD simulation results (filled symbols).
}
\label{Fig:2}
\end{figure*}

We next focus on the determination of the temperature of maximum density at 1 bar
for different molalities of perchloric acid.
A summary of the computed TMD values and the corresponding densities at the TMD is given in table~\ref{Table3}
for all concentrations considered in this work.
In figure~\ref{Fig:2}(a), 
we present the density as a function of temperature for four perchloric acid concentrations:
0.98 $m$, 0.69 $m$, 0.49 $m$, and 0.30 $m$. 
These systems correspond to simulation boxes containing 10, 7, 5, and 3
H$_3$O$^+$--ClO$_4^-$ ion pairs, respectively, 
in 555 H$_2$O molecules.
Specifically, the 0.98 $m$ system includes 10 H$_3$O$^+$ and 10~ClO$_4^-$ ions; 
the 0.69 $m$ system contains 7 H$_3$O$^+$ and 7 ClO$_4^-$ ions; 
and the 0.49 $m$ and 0.30 $m$ systems comprise 5 and 3 ion pairs, respectively, 
all solvated in 555 water molecules. 
Further details regarding the molecule concentrations are provided in reference~\cite{Gamez2025_H3O}.
To the best of our knowledge, experimental TMD
values for perchloric acid solutions are not available in the literature.
However, previous studies have shown that the Madrid-2019 force field reproduces experimental TMDs
and densities at the TMD for a broad set of aqueous electrolyte systems with good
accuracy~\cite{sedano2022maximum,gamez2023building,TMDs_Blazquez2024}.
Specifically, for related perchlorate and oxonium salts, the extended Madrid-2019
model~\cite{Blazquez2025perchlorate,Gamez2025_H3O} predicted
densities at the TMD with deviations below 0.1$\%$ and TMD values within approximately
1$\%$ of experimental data.

On the other hand, a particularly interesting property is the lowering of the temperature of maximum
in density of water, i.e., the shift in the TMD defined as
$\Delta = \text{TMD}_{\text{solution}}  - \text{TMD}_{\text{water}}$.
For dilute aqueous solutions, this dependence follows the Despretz law~\cite{despretz1839,despretz1840},
which states that $\Delta$ varies linearly with the molality $m$ according to,
\begin{equation}
\Delta = K_m m,
\end{equation}
where $K_m$ is the Despretz constant expressed in molality units.
In this dilute regime, it is reasonable to assume that the solvent effectively
screens ion-ion interactions, and thus an additive group contribution approach is often employed
in the literature.
In this framework, $K_m$ can be decomposed into the individual ionic contributions
($K_m^{\pm}$)~\cite{sedano2022maximum} as follows,
\begin{equation}\label{eq1}
   K_m =  (\nu_+ K_m^+ + \nu_- K_m^-),
\end{equation}
where $\nu_+$ and $\nu_-$ are the stoichiometric coefficients of the cation and anion, respectively
(for instance, in the case of HClO$_4$, $\nu_+ = \nu_- = 1$).
The equation~\ref{eq1} provides a reliable approximation for dilute concentrations,
where ion-ion interactions play a minor role.
Since the Despretz law relates the molality of the solution to the shift in the TMD ($\Delta$),
it is particularly useful to express our results in this manner.
To evaluate $\Delta$ for the Madrid-2019 force field, we used the TMD value obtained with the
TIP4P/2005 water model reported in reference~\cite{sedano2022maximum}, namely TMD$_{\text{water}}=277.3$~K.
The calculated $\Delta$ values relative to pure water for each molality are listed in table~\ref{Table3}.
Briefly, we estimated the Despretz constant from the ionic contributions of the perchlorate anion,
$K_m^{\mathrm{ClO_4^-}} = -13.5$ K$\cdot$kg$\cdot$mol$^{-1}$~\cite{Blazquez2025perchlorate},
and the oxonium cation,
$K_m^{\mathrm{H_3O^+}}=-4.5$ K$\cdot$kg$\cdot$mol$^{-1}$~\cite{Gamez2025_H3O},
obtaining $K_m = -18$ K$\cdot$kg$\cdot$mol$^{-1}$.
For the 1 $m$ solution, the calculated TMD$_{\mathrm{HClO_4}}$ is 259.3 K.
The remaining TMD values at different concentrations are summarized in table~\ref{Table3},
together with those obtained from a third-order polynomial fit to the MD simulation data.
As shown in figure\ref{Fig:2}(b), $\Delta$ decreases linearly with increasing molality ($R^2 = 0.998$),
yielding a slope of $K_m = -21.093 $~K$\cdot$kg$\cdot$mol$^{-1}$ with a slight deviation
from the ionic contributions reported in references~\cite{Blazquez2025perchlorate,Gamez2025_H3O}.
Notice that to determine the TMD accurately runs of about
100--200~ns for each temperature (typically eight temperatures in total) were needed so that
each TMD requires more than one microsecond of simulation time.
In all TMD results presented in table~\ref{Table3}, the uncertainty in the TMD values is within 1 K.

\begin{table}[ht!]
\centering
\caption{
Results of the temperature of maximum density
and the density ($\rho_\mathrm{max}$) of the perchloric acid solutions
at different molalities.
}
\vspace{0.2cm}
\label{tab:TMD}
\begin{tabular}{
   >{\centering\arraybackslash}m{1.9cm}
   >{\centering\arraybackslash}m{3cm}
   >{\centering\arraybackslash}m{3cm}
   >{\centering\arraybackslash}m{3cm}
}
\hline
Molality ($m$)& TMD (K) Simulation &TMD (K) Calculated from references~\cite{Blazquez2025perchlorate,Gamez2025_H3O}&
$\rho_{\max}$ (kg/m$^3$) Simulation \\
\hline
  0.30 & 270.5 & 271.9 & 1019.8 \\
  0.49 & 266.5 & 268.5 & 1032.4 \\
  0.69 & 262.9 & 264.9 & 1044.5  \\
  0.98 & 256.9 & 259.7 & 1062.6  \\
\hline
\end{tabular}
\label{Table3}
\end{table}

{It is important to point out that the TMD of pure water is already well 
reproduced by the TIP4P/2005 model, as reported in several previous studies~\cite{Vega2005,vega2011}. 
The addition of salt to water shifts the TMD, and the magnitude of this shift depend
sensitively on the nature of the dissolved electrolyte. 
In particular, earlier work has shown that the TMD displacement is highly sensitive to the ionic
charge distribution, as discussed by Blazquez {et al.}~\cite{blazquez2023scaled}. 
They demonstrated that electrolyte models employing unit ionic charges
($\pm 1e$) lead to an incorrect prediction of the TMD shift.
In this context, the use of scaled ionic charges (typically in the range 0.80--0.85$e$)
is essential for a reliable description of the TMD in aqueous electrolyte solutions. 
While the absolute value of the TMD is primarily determined by the underlying water model, 
the presence of ions introduces a secondary but non-negligible perturbation. 
As a result, the TMD exhibits a negative and approximately linear shift with increasing salt 
concentration, in agreement with the Despretz law expected for dilute electrolyte solutions.
Finally, the TMD values reported here provide useful reference values for future experimental
work aimed at determining the temperatures of maximum density and the corresponding densities
across different concentrations.
}

\subsection{Radial distribution function}

Perchloric acid (HClO$_4$) consists of a central chlorine atom bonded to four oxygen atoms,
one of which is also covalently bonded to a hydrogen atom. When dissolved in water it yields a perchlorate anion and (after reacting with one molecule of water) an oxonium cation.
The structural characterization of perchloric acid in aqueous solution is of particular interest,
as the interactions between water molecules and the perchloric acid reveal specific solvation
patterns that can be analyzed through the radial distribution function.
However, experimental information regarding the microscopic structure of aqueous perchloric acid remains limited.

\begin{figure*}
\begin{center}
\includegraphics[height=4.5cm,clip]{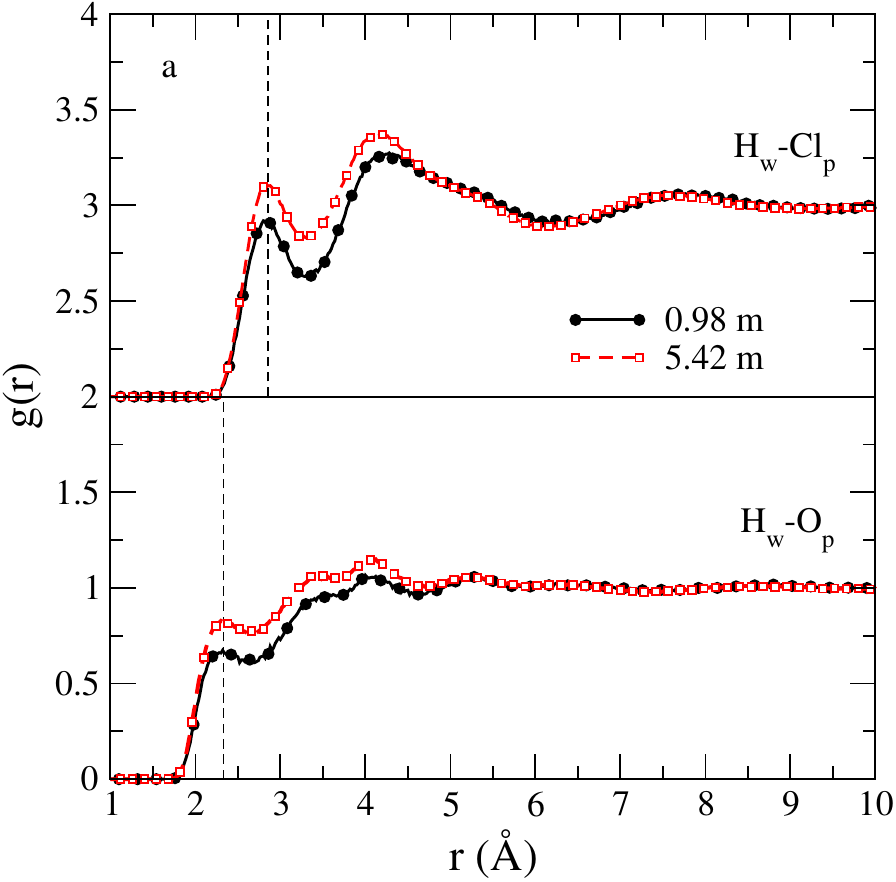}
\includegraphics[height=4.5cm,clip]{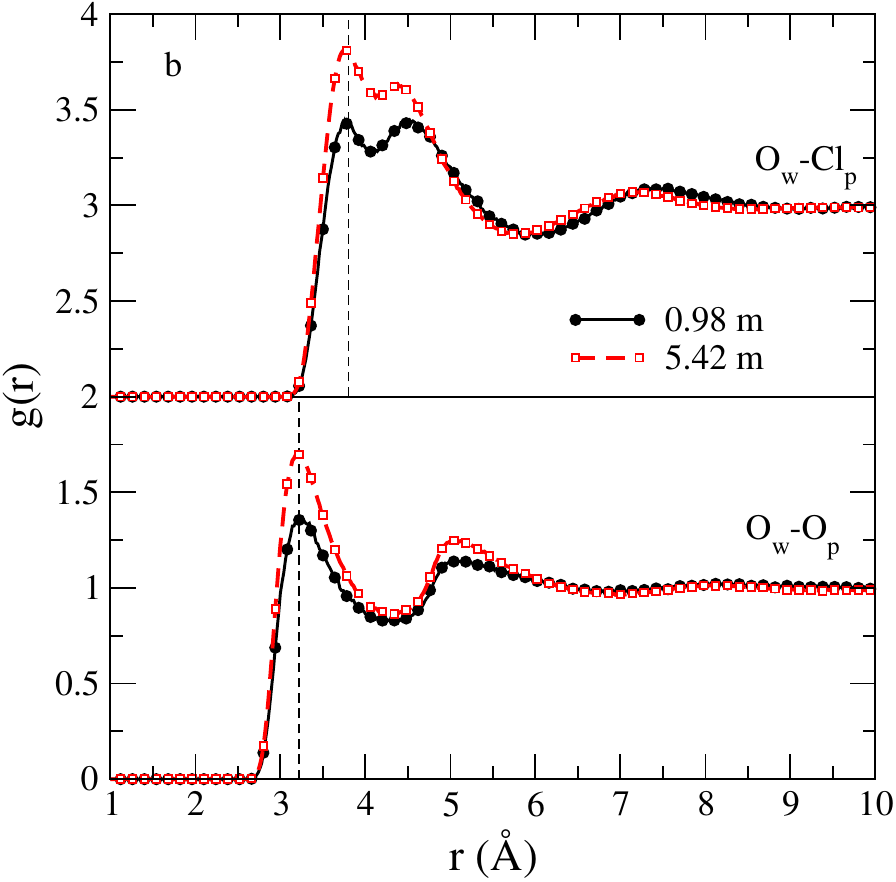}
\includegraphics[height=4.5cm,clip]{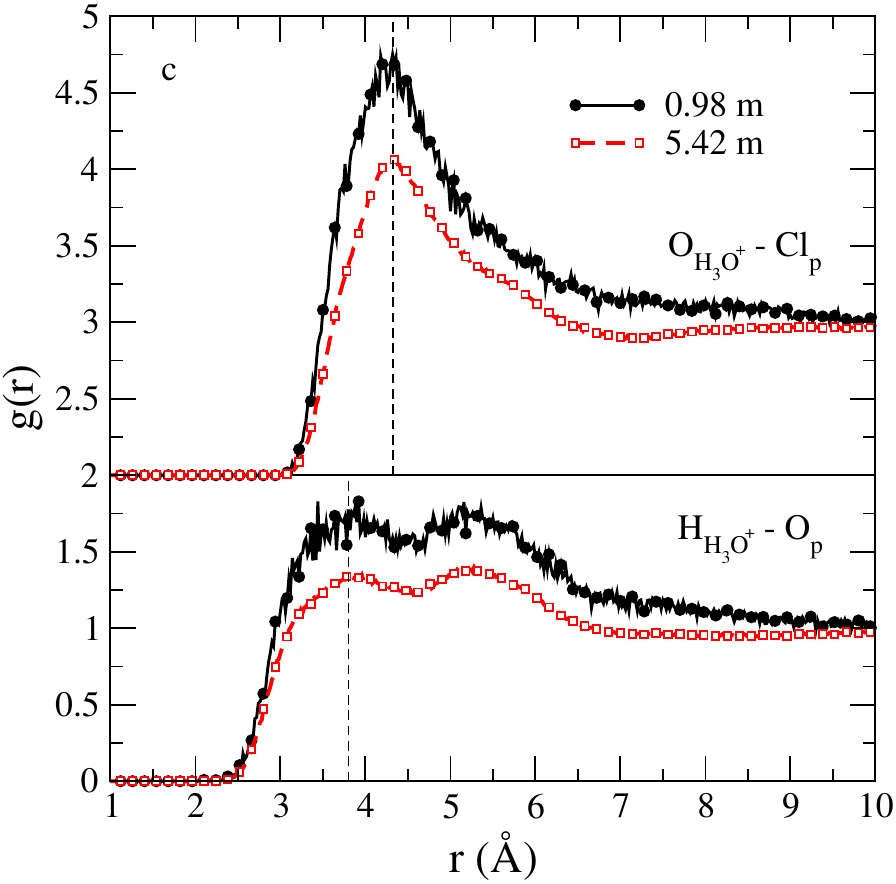}
\end{center}
\caption{(Colour online) 
Radial distribution functions for aqueous solutions of perchloric acid (HClO$_4$)
at low (0.98 $m$, solid black line) and high (5.42 $m$, dotted red line) concentrations at 298.15 K and 1~bar.
Panel (a): RDFs for H$_w$--Cl$_p$ and H$_w$--O$_p$ interactions.
Panel (b): RDFs for O$_w$--Cl$_p$ and O$_w$--O$_p$ interactions.
Panel (c): RDFs for O$_{\text{H}_3 \text{O}^+}$--Cl$_p$ and  H$_{\text{H}_3 \text{O}^+}$--O$_p$, interactions. The vertical dotted lines correspond to the experimental values for the position of the first maximum of the RDFs.
}
\label{Fig:3}
\end{figure*}

In figure~\ref{Fig:3}, the RDFs are shown for
panel~(a): H$_w$--Cl$_p$ and H$_w$--O$_p$,
panel~(b): O$_w$--Cl$_p$ and O$_w$--O$p$, and
panel~(c): O$_{\text{H}_3 \text{O}^+}$--Cl$_p$ and  H$_{\text{H}_3 \text{O}^+}$--O$_p$,
as obtained from MD simulations.
The RDFs are presented for two molalities, 0.98~$m$ and 5.42~$m$, to analyze how the
structure of the solution changes at low and at high concentration.
In figure~\ref{Fig:3}(a) (bottom panel), the H$_w$--O$_p$ radial distribution function is shown,
with the first peak of the RDF indicated by a dotted line.
The first peak appears at 2.3~\AA, which is approximately 0.5~\AA\ larger than the value
reported by Weber {et al.}~\cite{Weber2001}.
This difference probably arises because their study focused on microsolvated clusters,
whereas our simulations correspond to bulk aqueous solutions containing a much larger
number of water molecules.
The RDFs for O$_w$--Cl$_p$ (top panel) and O$_w$--O$_p$ (bottom panel) at both concentrations
are shown in figure~\ref{Fig:3}(b).
In the O$_w$--Cl$_p$ RDF shown in figure~\ref{Fig:3}(b) (top panel), two distinct maxima are observed.
The first peak at 3.77~\AA\ arises from the penetration of water molecules into the tetrahedral structure
of the perchlorate anion, while the second peak at 4.40~\AA\ corresponds to water molecules located outside
the tetrahedron. The position of the first maximum (3.77~\AA) is consistent with the range
of 3.6--3.8~\AA\ reported by Neilson~{et~al.}~\cite{Neilson1985}.
The O$_w$ $\cdots$ O$_p$ distance, which represents the separation between water
oxygen atoms and the oxygen of the perchlorate anion, exhibits a prominent first peak
at approximately 3.20~\AA~(see figure~\ref{Fig:3}(b) bottom panel).
This value is in good agreement with experimental data, which report a distance
of 3.07~\AA~from IR studies of ClO$_4^-$ hydration~\cite{PehrAake1991},
and values within the range 2.4--3.2~\AA~for the O$_w$ $\cdots$ O$_p$ separation~\cite{Neilson1985}.
In figure~\ref{Fig:3}(c) (top panel), the O$_{\text{H}_3 \text{O}^+}$--Cl$_p$ RDF is shown.
The absence of a well-defined first minimum in this RDF indicates that contact ion pairs (CIPs)
are not present under the simulated conditions.
For the H$_{\text{H}_3 \text{O}^+}$--O$_p$ distances (corresponding to the hydrogen atom of the oxonium
cation and the oxygen atom of the perchlorate anion, respectively), the first peak of the RDF appears
at 3.8~\AA\ (see figure~\ref{Fig:3}(c), bottom panel). By contrast, for the microsolvated cluster
HClO$_4$--(H$_2$O)$_3$ reported by Weber~{et al.}~\cite{Weber2001}, this distance is
approximately 1.5~\AA.
The difference arises because their system includes only three water molecules,
whereas our study considers a fully solvated aqueous solution.
Finally, to construct the perchlorate molecule, the intramolecular O$_p$--Cl$_p$ distance
was set to 1.43~\AA~\cite{Blazquez2025perchlorate}, which is consistent with the value obtained from neutron
diffraction measurements~\cite{Neilson1985}.
{To the best of our knowledge, experimental structural
data for HClO$_4$ in water are scarce. In this context, computational results provide valuable
microscopic insight into the solvation structure and ion-water correlations, and they can serve 
as a reference for future experimental investigations, including neutron diffraction studies 
that may emerge from different laboratories.
The existing experimental structural information is limited to perchlorate salts, such as sodium and
potassium perchlorate~\cite{lenton2017highly}, whose hydration environments differ from that of 
perchloric acid solutions. 
}

\subsection{Self-diffusion coefficients and shear viscosity}

\begin{figure*}
\begin{center}
\includegraphics[height=5.7cm,clip]{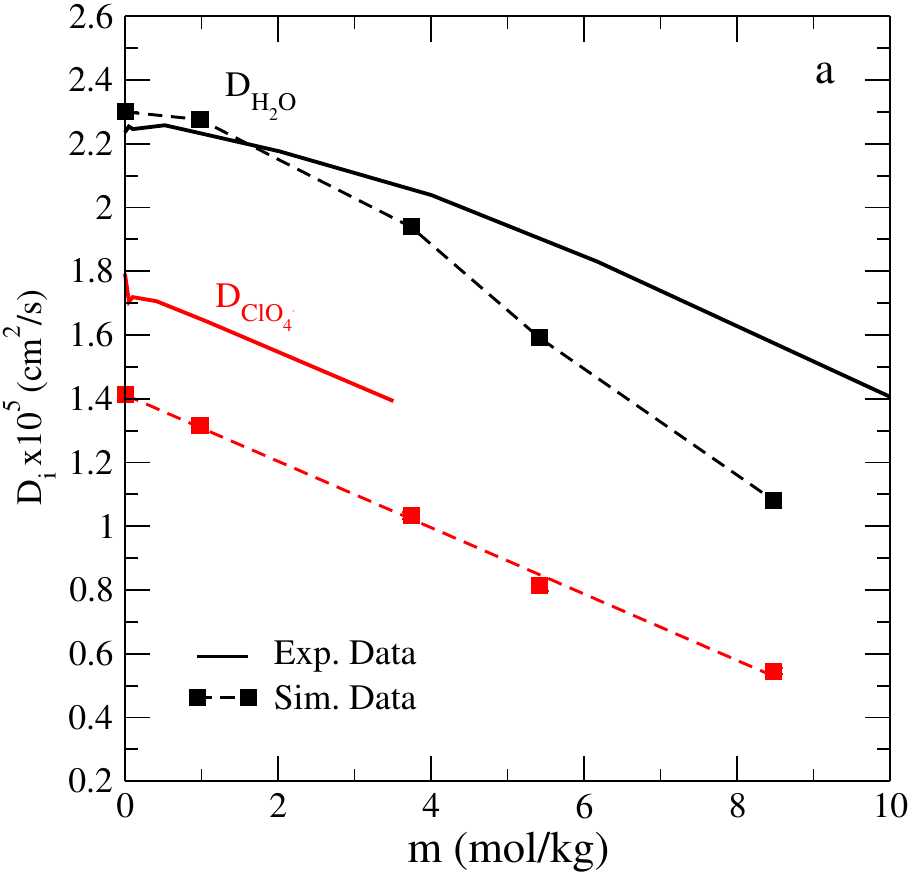}
\includegraphics[height=5.7cm,clip]{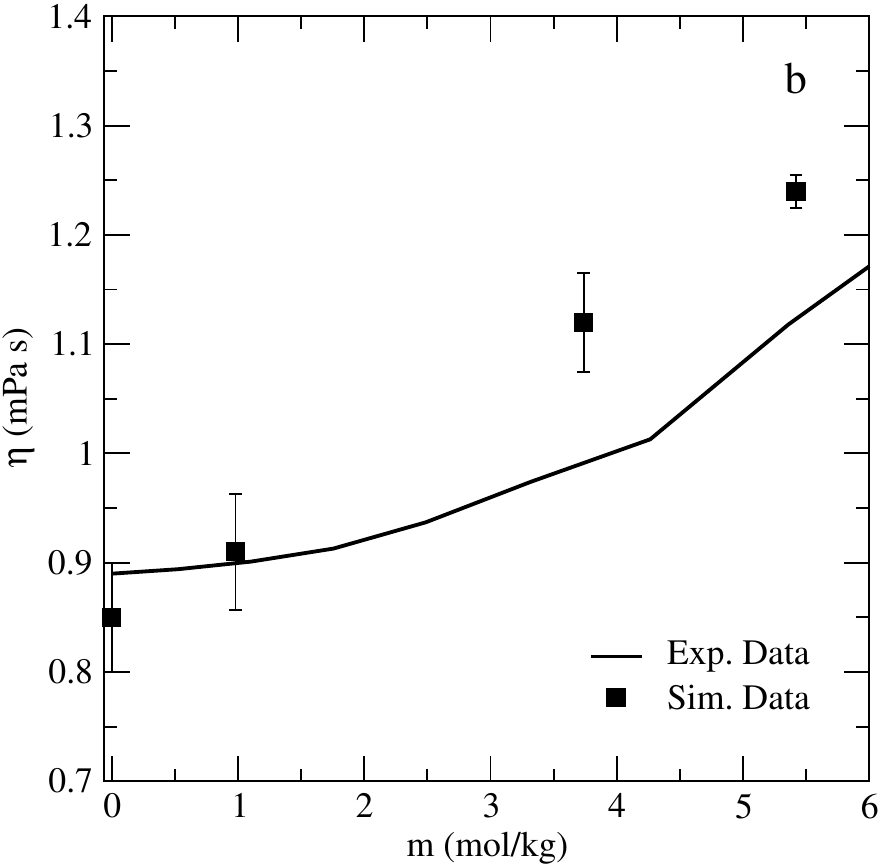}
\end{center}
\caption{(Colour online) 
Panel (a):
self-diffusion coefficients of water and perchloric ion (ClO$_4^{-}$) as a function of molality at
298.15 K and 1 bar. Filled squares represent simulation results, while dashed lines are included
as visual guides.
The experimental was obtained from reference~\cite{Malhotra1993}.
Panel (b):  shear viscosity as a function of molality for aqueous solutions of perchloric acid (HClO$_4$)
at 298.15~K and 1~bar.
Filled squares represent simulation results, while continuous lines correspond to experimental data
from reference~\cite{Brickwedde1949properties}.
}
\label{Fig:4}
\end{figure*}

Experimental values of self-diffusion coefficients ($D$) for individual ions in water
or for water in the presence of ions are scarce in the literature.
In figure~\ref{Fig:4}, we present a comparison between experimental data and MD
simulation results for the self-diffusion coefficients of water and the perchlorate ion
(ClO$_4^-$) as a function of molality at 298.15~K and 1~bar.
For comparison purposes, we employed the experimental data reported by Malhotra and
Lawrence~\cite{Malhotra1993} at 25\textcelsius~.
The perchlorate ion is known to be weakly solvated and does not readily form cation-anion
complexes, making experimental measurements of its self-diffusion coefficient feasible.
The diffusion coefficients of water, $D_{\text{H}_2\text{O}}$, obtained from our simulations
show very good agreement with experimental results at low molalities.
Here, the Yeh and Hummer finite-size corrections~\cite{Yehhummer} were applied to account
for the system-size dependence of the diffusion coefficients.
As the perchloric acid concentration increases, the simulated self-diffusion coefficients of
water exhibit larger deviations from experiment; however, the overall trend with increasing
concentration is well reproduced.
For the perchlorate ion (ClO$_4^-$), the MD simulations also capture the general trend of the
experimental data with qualitative agreement.
The simulated diffusion coefficients reproduce the correct dependence on concentration,
confirming that the extended Madrid-2019 force field provides a reasonable description
of the dynamical behavior of this system.

In reference~\cite{mul:jpc96} the experimental self-diffusion coefficient of pure water is reported as
$2.3\times10^{-5}$~cm$^2$/s, while Tanaka~\cite{tanaka1975measurements}
measured a value close to $2.2\times10^{-5}$~cm$^2$/s.
In our MD simulations, we obtained a corrected diffusion coefficient of $2.3\times10^{-5}$~cm$^2$/s
for water using the TIP4P/2005 model, in excellent agreement with both experimental results.
The self-diffusion coefficients of the perchlorate ion (ClO$4^-$) at 25~\textcelsius~and infinite dilution
are well documented in the literature. From the limiting ionic conductance and using the
Nernst-Einstein relation, a value of
$D^{\infty,\text{exp}}_{\text{ClO}_4^-}=1.792\times10^{-5}$~cm$^2$/s
has been reported~\cite{Stefan1995}. In this work, we estimated
$D^{\infty,\text{sim}}_{\text{ClO}_4^-}$
by linearly extrapolating the simulated diffusion coefficients of ClO$4^-$ to the limit $m\rightarrow 0$,
obtaining $D^{\infty,\text{sim}}_{\text{ClO}_4^-}=1.412\times10^{-5}$~cm$^2$/s.
This value is slightly lower than the experimental result,
as shown by the red dotted and continuous lines in figure~\ref{Fig:4}(a).

Finally, after analyzing bulk densities, temperatures of maximum density, structural features,
and self-diffusion coefficients, the estimation of shear viscosity provides an additional stringent
test of the performance of the extended Madrid-2019 force field in modelling the dynamic properties
of perchloric acid solutions.
It is important to note that viscosity calculations are computationally demanding;
therefore, we restricted our analysis to concentrations between 0 and 6 $m$.
Figure~\ref{Fig:4}(b) shows the shear viscosity as a function of molality for aqueous perchloric acid
(HClO$_4$) solutions at 298.15~K and 1~bar.
{We have included a representative running integral of the viscosity
in the Supplementary material for the system at three considered concentrations
(see figure~S3).
}
The agreement between simulation and experiments is quite good except at the higher concentration
considered where the model slightly   
{overestimates  the experimental result.
At low concentrations a good prediction of the viscosity is ensured as the TIP4P/2005 model of water nicely reproduces the experimental viscosity of pure water.
Blazquez~{et~al.}~\cite{Blazquez2025perchlorate} reported that, in the case of sodium perchlorate
(NaClO$_4$) at 298.15 K, the shear viscosity tends to be overestimated throughout the concentration 
range and the same behavior is found here. 
Interestingly the deviation from experiment found here at high concentrations is smaller than the
one found in the previous work for NaCl and other simple electrolytes. 
Although the use of the scaled global charge $q = \pm0.85e$ is capable of reproducing  
many experimental properties across the entire range of molalities studied it tends to overestimate 
the viscosities. 
Certainly, the agreement between the model and the experimental transport properties can be improved 
by reducing the value of the scaled charge to $q = \pm0.75e$ as discussed in recent 
works \cite{Habibi2022,blazquez2023scaled} and/or modifying the internal charge distribution of the 
polyatomic ions as has been shown recently for the perchlorate anion\cite{Blazquez2025perchlorate}. 
In any case, the predictions of the viscosities presented in  figure~\ref{Fig:4}(b)  
are already quite reasonable.}

\section{Conclusions and perspectives}\label{Conclusions}

We have applied the extended Madrid-2019 force field to model perchloric acid (HClO$_4$)
in TIP4P/2005 water, using the published parameters for the perchlorate anion (ClO$_4^-$) and
the oxonium cation (H$_3$O$^+$) reported in references~\cite{Blazquez2025perchlorate,Gamez2025_H3O}.
Notably, the perchloric acid model was built solely using these published parameters without
retuning cross-interaction parameters and using LB combining rules for the anion-cation interactions.
The model reproduces a wide set of properties with good quantitative accuracy:
density --- molality (including trends up to 10 $m$), temperatures of maximum density,
ion-ion and ion-solvent structural features (RDFs and coordination), self-diffusion coefficients,
and shear viscosities (within the concentration ranges studied).
These results indicate that the extended Madrid-2019 parametrization provides a reliable and broadly
applicable description of perchloric acid solutions.
As expected, transport properties remain most sensitive to the ionic charge model.
Using the scaled-charge model with a net charge of $\pm$0.85$e$
yields viscosities in reasonable agreement with experiment data across the studied molality range;
although both prior work and our present results suggest that transport properties
(and diffusion at infinite dilution) can be further improved by adjusting the ionic
charge magnitude or the intramolecular partial-charge distribution.
For example, the Madrid-Transport model
with global ion charge by $q = \pm0.75e$ has been shown to be in better agreement with the experimental data
for some transport properties~\cite{blazquez2023scaled}.
In perspective, the force field employed here combines a relative simplicity with good predictive
capability for thermodynamic, structural, and dynamical properties of perchloric acid solutions.\\

\section*{Dedications}

We are deeply honored to dedicate this work to the memory of 
Prof. Stefan Soko{\l}owski, in recognition of his distinguished scientific career 
and his invaluable contributions as a researcher, mentor, and professor.
Stefan made a profound impact on the field of adsorption theory through
the use of density functional approaches and molecular dynamics simulations.
His research spanned a wide range of topics, including the phase behavior of
complex fluids and the effects of confinement on fluid properties.

He began his scientific journey by applying virial expansions to nonuniform
model fluids and by developing new methodologies based on integral equations,
Monte Carlo simulations, and theoretical extensions for non-spherical
molecules and multicomponent mixtures.
Later, Stefan held a postdoctoral position under the supervision of Prof. W.A. Steele
at Pennsylvania State University, where he further developed his ideas in
the area of density functional theory and molecular dynamics simulations to
explore a broader range of problems, including the phase behavior of fluids
 under external fields.
In 1988, Soko{\l}owski was awarded an Alexander von Humboldt Fellowship at the
Ruhr-Universit{\"a}t Bochum, where he worked very successfully with
Prof. Johan Fischer.
It was in 1989 when one of us (C.V.) met Stefan for the first time while living in Bochum.
Stefan was an example for C.V. about hard work, honesty, brightness and nice personality
can be together in the same person. Later in 2005 C.V. spent one of the best summers of his
life in Lublin, Poland, thanks to the hospitality of Stefan (including the parties in his house).
Stefan and Johan Fischer together extended several formulations of
density functional theory and Born--Green--Yvon integral equations to the
study of mixtures in pores and carried out molecular dynamics simulations
of wetting transitions at the argon-carbon dioxide interface.
To contribute to the study of transport phenomena in granular materials,
Soko{\l}owski joined Forschungszentrum J{\"u}lich as a visiting professor
(1990-1991), where he co-authored one of the most cited Physical Review Letters
papers on molecular dynamics simulations of vibrated granular systems.
Later, he was invited by Prof. Doug Henderson to spend a year as a visiting
scientist at the Universidad Aut\'onoma Metropolitana - Iztapalapa (UAM-Izt)
in Mexico City, where he became an influential collaborator and mentor.
It was in 2018 when two of us (M.C.S. and V.M.T.) met Stefan during one of his many
research visits to the Institute of Chemistry at the Universidad Nacional Aut\'onoma de M\'exico (UNAM)
in Mexico City. We remember Stefan as a brilliant scientist, full of fresh and innovative ideas.
Under his guidance, and in collaboration with Professor O. Pizio, we published a couple of articles
on density functional theory and molecular dynamics simulations. We fondly recall enjoying excellent
cups of cognac with coffee, savoring authentic Polish food, and sharing lively discussions
about science and music in Stefan's company.

Stefan was a brilliant and rigorous scientist, capable of making complex
concepts accessible and inspiring through his teaching and mentorship.
His guidance profoundly influenced many respected researchers in both
Poland and Mexico. He was also a true colleague, generous, respectful,
and always willing to share his time, knowledge, and ideas.
We are honored to contribute to this special issue in memory of Stefan,
celebrating his life, his achievements, and his enduring legacy.
His passing is a great loss to the scientific community.
We will always remember Stefan as a generous man full of wisdom and
curiosity, ever eager to share it with his friends and colleagues.
Thank you, Stefan, for so much.\\

\section*{Supplementary material}

In the Supplementary material, we compile the numerical (raw data) and graphical information
of the simulation results of perchloric acid considered in this work for the following properties:
(\emph{i}) simulation results for the density ($\rho^{\rm sim}$) as a function of the molality ($m$);
(\emph{ii}) simulation results for the density ($\rho^{\rm sim}$) as a function of the temperature ($T$);
(\emph{iii}) simulation results for the shift in the TMD as a function of the molality ($m$);
(\emph{iv}) simulation results for the self-diffusion coefficient ($D_i^{\rm sim}$) as a function of the molality ($m$);
(\emph{v}) simulation results for the viscosity ($\eta^{\rm sim}$) as a function of the molality ($m$).
We also include a figure showing the running values of the viscosity as a function of the correlation
time and molality variation at 298.15 K and 1 bar.
Additionally, we have included the molecular dynamics input files, \emph{topol.top} and \emph{conf.g96},
for reference and reproducibility purposes.

\section*{Acknowledgements}

V.M.T. and M.C.S. acknowledge the financial support from SECIHTI-Mexico through the program
``Convocatoria Ciencia B\'asica y de Frontera 2023--2024'' grant number CBF2023-2024-2725,
Proyecto PEAPDI2025 ``Modelado Molecular de Disoluciones Electrol\'iticas con Disolvente Agua''
and the Department of Physical Chemistry at the Universidad Complutense de Madrid for its hospitality.
C.V and S.B. aknowledge funding from project PID2022-136919NB-C31 of the Spanish Ministry of Science and Innovation.


\bibliographystyle{cmpj} 
\bibliography{references}

\ukrainianpart

\title{	Молекулярно-динамічне дослідження перхлоратної кислоти з використанням розширеного силового поля Мадрид-2019}
%
%
\author{
	M. Круз-Санчес\refaddr{label1},
	С. Бласкес\refaddr{label2},
	C. Вега\refaddr{label2},
	В. M. Трехос\refaddr{label1}
}
\addresses{
	\addr{label1} Факультет хімії, Столичний автономний університет Ізтапалапа,
	просп. Сан Рафаель Атлікско 186, Віцентіна, 09340, Мехіко, Мексика
	\addr{label2} Кафедра фізичної хімії, Університет Комплутенсе в Мадриді, 28040 Мадрид, Іспанія
}

\makeukrtitle

\begin{abstract}
	\tolerance=3000%
	Перхлоратна кислота (HClO$_4$) широко використовується для отримання перхлоратних солей, що застосовуються в паливних матеріалах, промисловості, хімії навколишнього середовища та біології.
	У цій роботі ми використовували міжмолекулярні параметри з розширеного силового поля Мадрид-2019 для перхлорат-аніона (ClO$_4^-$) та оксонієвого катіона (H$_3$O$^+$) разом з водою TIP4P/2005 для моделювання розчинів перхлоратної кислоти.
	Силове поле використовує ефективні заряди $\pm0.85e$ для одновалентних іонів і широко застосовується для водних іонних систем.
	Ми використовували цю модель для прогнозування термодинамічних властивостей [густини та температури максимуму густини (TMD)], структурних особливостей (кореляції іон-вода, зокрема іон-водень та іон-кисень), а також транспортних характеристик (коефіцієнти самодифузії та в'язкість) розчинів перхлоратної кислоти при кількох концентраціях. Експериментальні густини прогнозуються з чудовою узгодженістю до 10~$m$.
	Ми також провели молекулярне моделювання в широкому діапазоні температур, щоб визначити термодинамічний дисперсійний розподіл  перхлоратної кислоти при різних молярностях. Прогнозовані значення в'язкості при температурі 298.15~K та тиску 1 бар добре узгоджуються з експериментальними даними для концентрацій нижче 4~$m$. Результати обговорюються з точки зору переваг та обмежень моделі.
	\keywords перхлоратна кислота, Мадрид-2019, молекулярна динаміка, силове поле
	
\end{abstract}

\lastpage
\end{document}



\title{SUPPLEMENTARY MATERIAL\\[0.7cm]
\textquotedblleft
Molecular Dynamics Study of Perchloric Acid Using the \\ Extended Madrid-2019 Force Field.\textquotedblright}

\author{Mario Cruz-Sánchez$^{1}$}
\author{Samuel Blazquez$^{2}$}
\author{Carlos Vega$^{2}$}
\author{Víctor M. Trejos$^{1,*}$}

\affiliation{$^{1}$Departamento de Química, Universidad Autónoma Metropolitana-Iztapalapa,
Av. San Rafael Atlixco 186, Col. Vicentina, 09340, Ciudad de México, México.}

\affiliation{$^{2}$Departamento de Química Física, Universidad Complutense de Madrid, 28040 Madrid, Spain.}

\email{vtrejos@izt.uam.mx}

\maketitle

\newpage

The Supplementary Material for the publication ‘Molecular Dynamics Study of Perchloric Acid Using
the Extended Madrid-2019 Force Field’ contains the compilation of the numerical
(raw data) and graphical information of the simulation results of perchloride acid solutions
considered in the main body of this work for the following properties:

\begin{itemize}
\item Simulation results for the density ($\rho^{\text{sim}}$) as a function of the molality
(\textit{m}) and the temperatures 283.15, 298.15, and 323.15 K.
The experimental data ($\rho^{\text{exp}}$) were obtained from Ref.~\cite{Brickwedde1949properties}.

\item Density as a function of temperature for aqueous solutions of perchloric acid (HClO$_4$)
at 0.98 (\textit{m}), 0.69 (\textit{m}),  0.49 (\textit{m}), and 0.30 (\textit{m}) at 1 bar.

\item Variation of $\Delta$ as a function of $m$ for perchloric acid solutions.

\item Self-diffusion coefficients of water (H$_2$O) and perchloric ion (ClO$_4^{-}$)
as a function of molality at 298.15 K and 1 bar.

\item Shear viscosity as a function of molality for aqueous solutions of perchloric acid (HClO$_4$)
at 298.15 K and 1 bar compared with experimental data obtained from Ref.~\cite{Brickwedde1949properties}.


\end{itemize}

\newpage
\section{Raw data}

\subsection{Bulk densities}

\begin{table}[H]
\centering
\caption{
Comparison between simulation results and experimental data for the bulk density
for perchloric acid solutions at 283.15 K, 298.15 K, and 323.15 K.
Experimental data were obtained from Ref.~\cite{Brickwedde1949properties}.
The relative percentage deviation is given by 
dev.(\%)$ = 100 \times | \rho^{\text{exp}} - \rho^{\text{sim}}|/ \rho^{\text{exp}} $.
Units: mol/kg for $m$ and kg/m$^3$ for density, $\rho$.
}
\scalebox{1}{
\begin{tabular}{|c|c|c|c|c|c|c|c|c|c|c|c|}
\hline  
\multicolumn{4}{|c|}{283.15 K} &
\multicolumn{4}{c|}{298.15 K} &
\multicolumn{4}{c|}{323.15 K} \\
\hline
$m$ & $\rho^{\exp}$ & $\rho^{sim}$ & dev.(\%) &
$m$ & $\rho^{\exp}$ & $\rho^{sim}$ & dev.(\%) &
$m$ & $\rho^{\exp}$ & $\rho^{sim}$ & dev.(\%) \\
\hline
0.00 & 1000.00 & 1001.16 & 0.12 &
0.00 & 997.10 & 997.30 & 0.02 &
0.00 & 988.00 & 988.33 & 0.03 \\
0.98 & 1054.05 & 1058.20 & 0.39 &
0.98 & 1049.41 & 1052.63 & 0.31 &
0.98 & 1037.69 & 1040.04 & 0.23 \\
1.93 & 1103.53 & 1108.53 & 0.45 &
1.93 & 1095.94 & 1100.83 & 0.45 &
1.93 & 1081.80 & 1085.56 & 0.35 \\
3.74 & 1187.25 & 1192.90 & 0.48 &
3.74 & 1177.90 & 1182.35 & 0.38 &
3.74 & 1160.29 & 1163.27 & 0.26 \\
5.42 & 1257.23 & 1262.06 & 0.38 &
5.42 & 1245.63 & 1247.43 & 0.14 &
5.42 & 1226.14 & 1227.79 & 0.13 \\
6.22 & 1287.30 & 1292.13 & 0.38 &
6.22 & 1275.70 & 1278.74 & 0.24 &
6.22 & 1254.99 & 1255.70 & 0.06 \\
7.00 & 1315.72 & 1319.60 & 0.29 &
7.00 & 1303.20 & 1305.61 & 0.18 &
7.00 & 1282.06 & 1281.85 & 0.02 \\
8.48 & 1366.94 & 1369.31 & 0.17 &
8.48 & 1353.39 & 1353.90 & 0.04 &
8.48 & 1330.20 & 1328.94 & 0.09 \\
\hline
\end{tabular}
}
\end{table}

\begin{table}[H]
\centering
\caption{
Density of perchloric acid solutions as a function of temperature
and molalities: 0.30 $m$, 0.49 $m$, 0.69 $m$, and 0.98 $m$.
Units: mol/kg for $m$, K for $T$, and  kg/m$^3$ for $\rho$.
}
\scalebox{1}{
\begin{tabular}{|c|c|c|c|c|c|c|c|}
\hline
\multicolumn{2}{|c|}{0.30 m} &
\multicolumn{2}{c|}{0.49 m} &
\multicolumn{2}{c|}{0.69 m} &
\multicolumn{2}{c|}{0.98 m} \\
\hline
$T$ & $\rho^{sim}$ &
$T$ & $\rho^{sim}$ &
$T$ & $\rho^{sim}$ &
$T$ & $\rho^{sim}$ \\
\hline
240 & 1010.23 & 240 & 1025.04 & 240 & 1039.39 & 230 & 1055.97 \\
245 & 1013.39 & 245 & 1027.75 & 245 & 1041.51 & 235 & 1058.30 \\
250 & 1015.80 & 250 & 1029.68 & 250 & 1043.03 & 240 & 1060.20 \\
255 & 1017.52 & 255 & 1031.05 & 255 & 1044.07 & 245 & 1061.33 \\
260 & 1018.66 & 260 & 1031.84 & 260 & 1044.34 & 250 & 1062.23 \\
265 & 1019.52 & 265 & 1032.11 & 265 & 1044.37 & 255 & 1062.54 \\
270 & 1019.65 & 270 & 1032.11 & 270 & 1044.12 & 260 & 1062.53 \\
275 & 1019.50 & 275 & 1031.66 & 275 & 1043.43 & 265 & 1062.06 \\
280 & 1019.14 & 280 & 1030.94 & 280 & 1042.44 & 270 & 1061.44 \\
285 & 1018.29 & 285 & 1029.91 & 285 & 1041.27 & 275 & 1060.43 \\
290 & 1017.29 & 290 & 1028.64 & 290 & 1039.79 & 280 & 1059.15 \\
 & & & & & & 285 & 1057.63 \\
 & & & & & & 290 & 1055.87 \\
\hline
\end{tabular}
}
\end{table}

\begin{table}[H]
\centering
\caption{
$\Delta$ of TMD for different molalities. 
Units: mol/kg for $m$ and K for $\Delta$.
}
\begin{tabular}{|c|c|}
\hline
$m$ & $\Delta$ \\
\hline
0.30 & -6.79 \\
0.49 & -10.84 \\
0.69 & -14.41 \\
0.98 & -20.38 \\
\hline
\end{tabular}
\end{table}

\newpage
\subsection{Self-Diffusion Coefficient}

\begin{table}[H]
\centering
\caption{
Simulation results for self-diffusion coefficient variation with molality for water, $D_{H_2O}$, 
and perchlorate anion, $D_{ClO_4^-}$.
Units in mol/kg for $m$ and cm$^2$/s for self-diffusion coefficient $D_i$.
}
\begin{tabular}{|c|c|c|}
\hline
$m$ & $D_{H_2O}$ & $D_{ClO_4^-}$ \\
\hline
0.00 & 2.300 & 1.412$^*$ \\
0.98 & 2.276 & 1.316 \\
3.74 & 1.939 & 1.035 \\
5.42 & 1.591 & 0.815 \\
8.48 & 1.081 & 0.545 \\
\hline
\end{tabular}

{\raggedright $^*$ Denotes linear interpolation from simulation data.\par}
\end{table}

\subsection{Shear viscosity}

\begin{table}[H]
\centering
\caption{
Comparison between simulation results and experimental data for shear viscosity 
variation with molality for HClO$_4$. Experimental data were obtained from Ref.\cite{Brickwedde1949properties}.
Units in mol/kg for $m$ and mPa$\cdot$s for shear viscosity $\eta$.
}
\begin{tabular}{|c|c|}
\hline
$m$ & $\eta^{sim}$ \\ 
\hline
0.00 & 0.85 \\
0.98 & 0.91 \\
3.74 & 1.12 \\
5.42 & 1.24 \\
\hline
\end{tabular}
\end{table}

\newpage

\section{Figures}
\subsection{Bulk densities}
\begin{figure}[H]
\centering
\includegraphics[height=7cm]{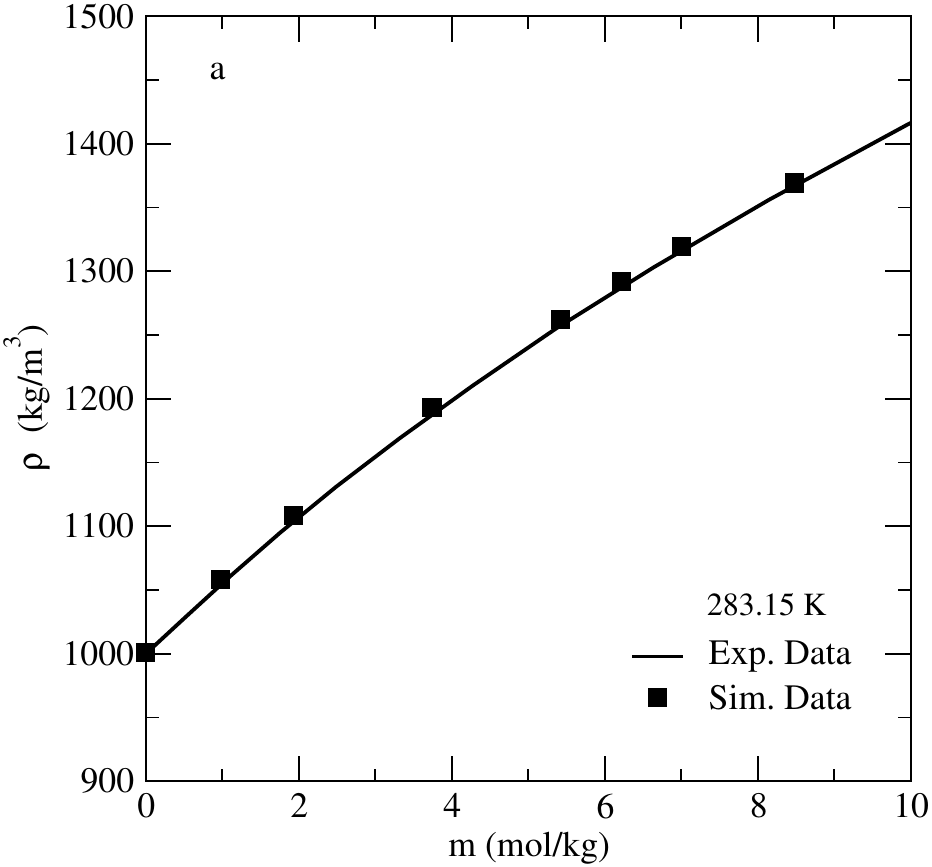}
\includegraphics[height=7cm]{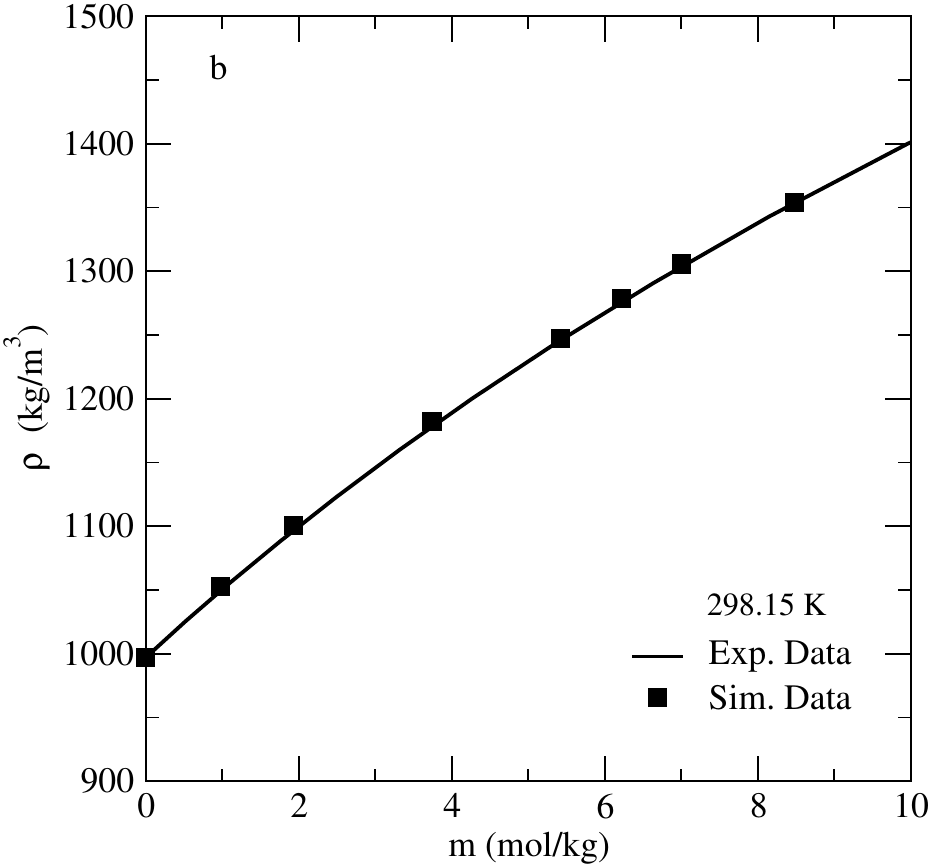}
\includegraphics[height=7cm]{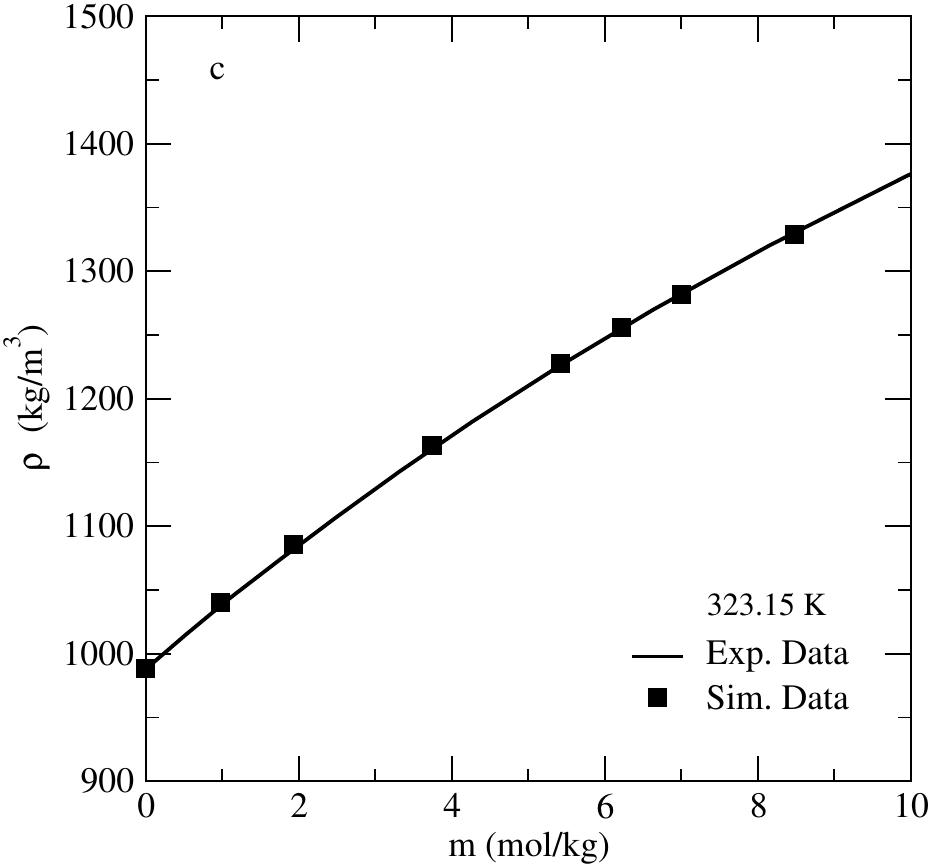}
\caption{
Density as a function of molality for aqueous HClO$_4$ solutions at 283.15 K (a),
298.15 K (b), and 323.15 K (c).
MD results: black squares; experimental data: continuous lines
(Ref.~\cite{Brickwedde1949properties}).
}
\end{figure}

\begin{figure}[H]
\centering
\includegraphics[height=7cm]{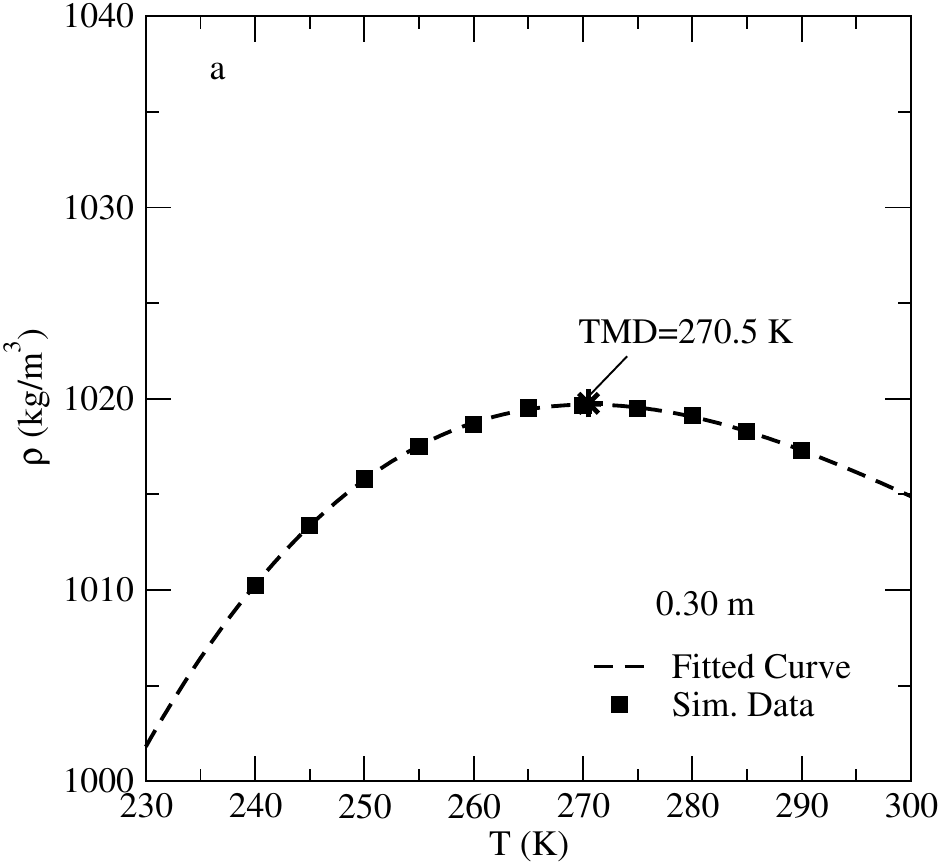}
\includegraphics[height=7cm]{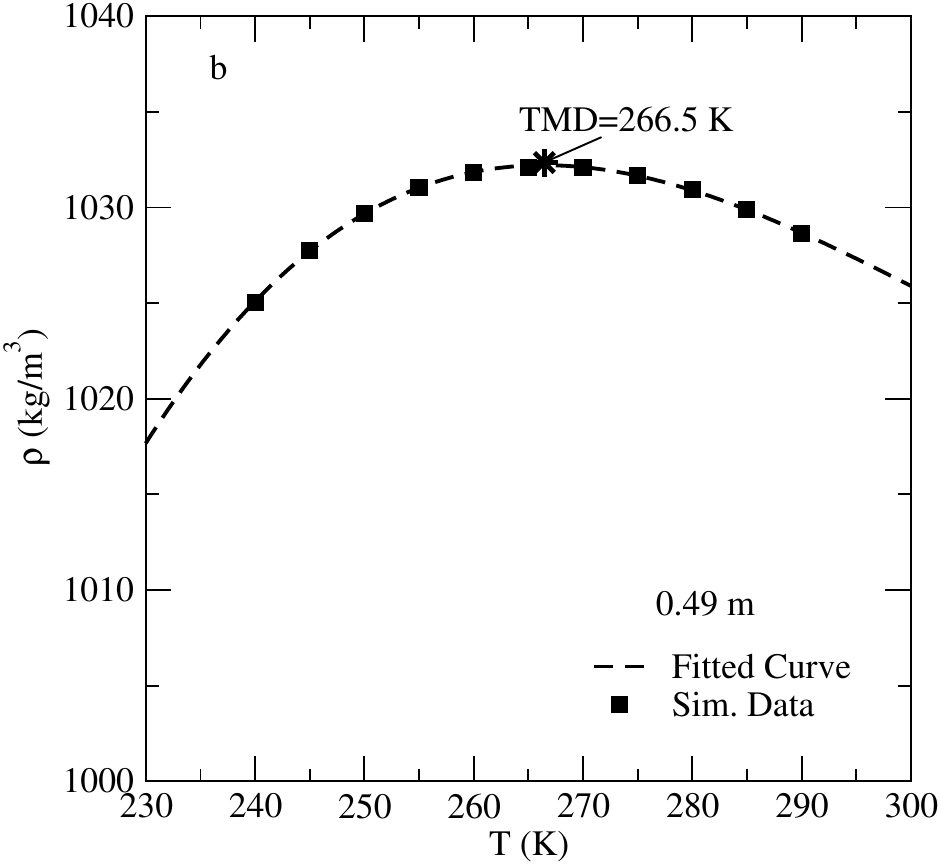}
\includegraphics[height=7cm]{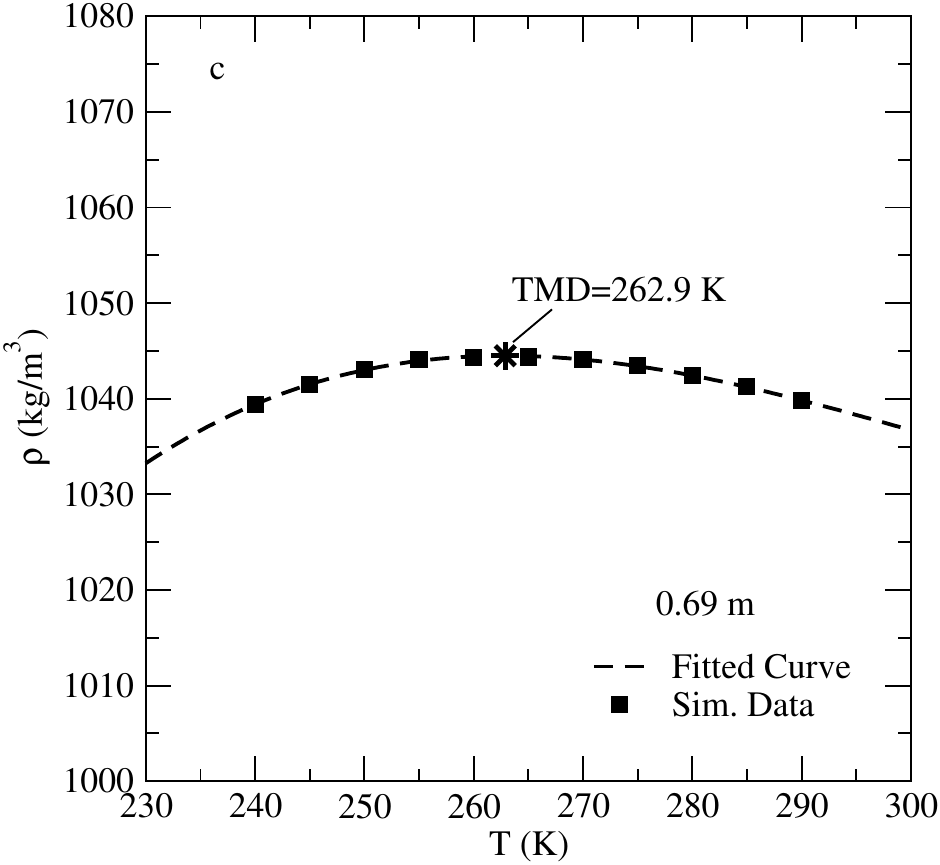}
\includegraphics[height=7cm]{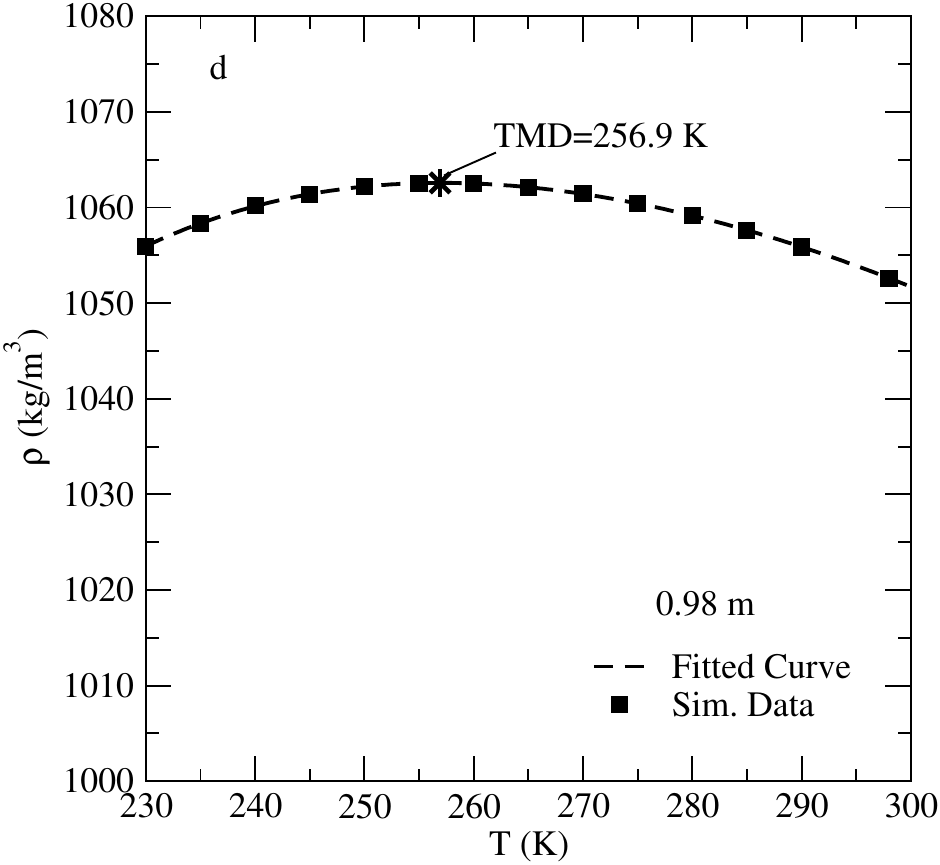}
\caption{
Density as a function of temperature at 0.30 m (a), 0.49 m (b), 0.69 m (c), 
0.98 m (d) molalities.
}
\end{figure}

\subsection{Shear viscosity}
\begin{figure}[H]
\centering
\includegraphics[height=8cm]{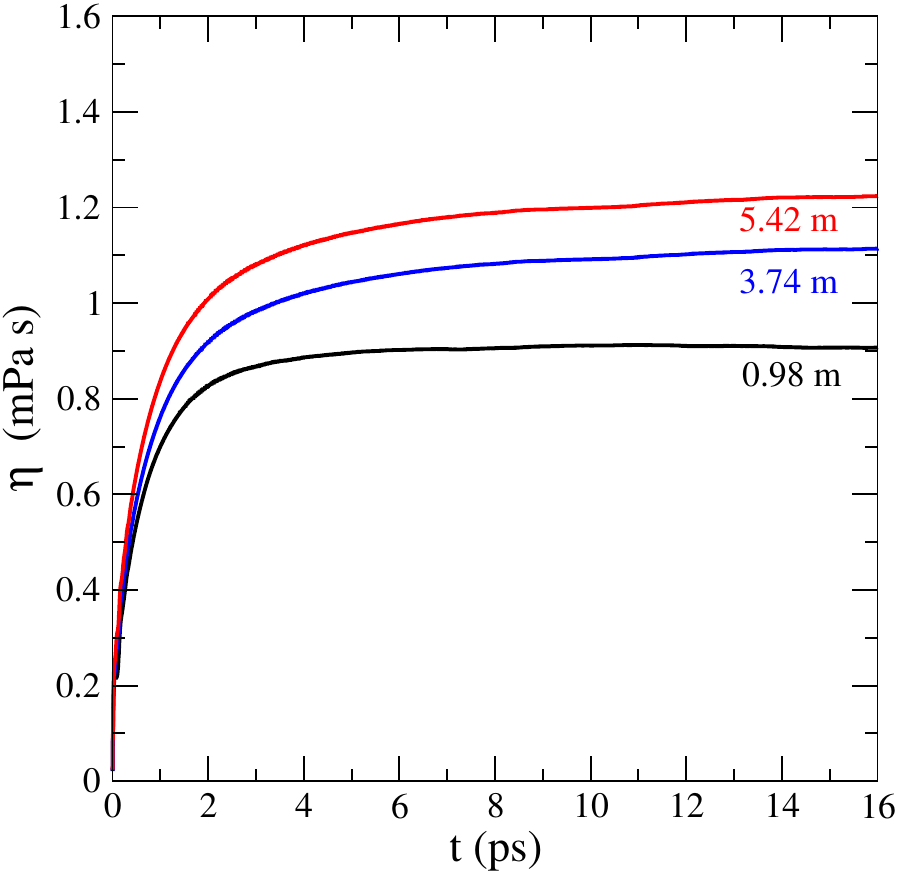}
\caption{
Running viscosity as a function of correlation time at 298.15 K and 1 bar.
}
\end{figure}


\bibliographystyle{ieeetr}
\bibliography{references_sup.bib}